\documentclass{LMCS}

\usepackage{amssymb}
\usepackage{bussproofs} 
\usepackage{pst-tree}
\usepackage[defblank]{paralist}
\usepackage{picins}
\usepackage{fancybox}
\usepackage{manfnt}
\usepackage{pstring}
\usepackage{mathrsfs}
\usepackage{hyperref,enumerate}
\definecolor{darkGreen}{rgb}{0.03,0.35,0.05}
\pstrSetArrowColor{darkGreen}

\newlength{\mylength}
\newenvironment{FramedTable}%
{\begin{table}[htbp]
\begin{Sbox}%
\setlength{\mylength}{\textwidth}%
\addtolength{\mylength}{-4\fboxsep}%
\addtolength{\mylength}{-2\fboxrule}%
\begin{minipage}{\mylength}}%
{\end{minipage}\end{Sbox}\shadowbox{\TheSbox}\end{table}}%

\newcommand\ie{{\it i.e.}} 
\newcommand\eg{{\it e.g.}} 

\makeatletter
\newcommand\defname[2][]{{\bf\em #2}%
  \edef\options{#1}%
  \ifx\options\@empty%
   \index{#2}%
  \else%
   \index{#1}%
  \fi%
}
\makeatother

\newcommand\betared{\rightarrow_\beta}
\newcommand\betasred{\rightarrow_{\beta_s}}
\newcommand\betaredtr{\twoheadrightarrow_\beta} 
\newcommand\subst[2]{\left[ #1/#2 \right]}
\newcommand\captsubst[2]{\{#1/#2 \}}

\newcommand{\elnf}[1]{\lceil #1\rceil} 
\newcommand\travset{\mathcal{T}rv}

\newcommand\dps{\displaystyle}
\newcommand\rulef[2]{\frac{\dps #1}{\dps #2}}

\newcommand{\ord}{\mathop{\mathrm{ord}}}
\newcommand{\rulename}[1]{\mathbf{({\sf #1})}}
\newcommand{\rulenamet}[1]{${\sf (#1)}$}
\newcommand{\encode}[1]{\overline{#1}}

\newcommand\typear{\rightarrow}

\newcommand{\makeset}[1]{\{\,{#1}\,\}}

\newcommand\union{\cup}

\newcommand\nat{\mathbb{N}}

\newcommand{\theroot}{\circledast} 
\newcommand{\sem}[1]{{[\![ #1 ]\!]}}
\newcommand{\revsem}[1]{{\langle\!\langle #1 \rangle\!\rangle}}
\newcommand{\syntrevsem}[1]{{\langle\!\langle #1 \rangle\!\rangle}_{\sf s}}

\newcommand{\oview}[1]{\llcorner #1 \lrcorner}
\newcommand{\pview}[1]{\ulcorner #1 \urcorner}
\newcommand{\filter}{\upharpoonright}

\newcommand\imp{\Longrightarrow}
\newcommand\zor{\vee}
\newcommand\zand{\wedge}

\newcommand\ialgol{{IA}}
\newcommand\iacom{\texttt{com}}
\newcommand\iaexp{\texttt{exp}}
\newcommand\iavar{\texttt{var}}
\newcommand\pcf{{PCF}}
\newcommand\ycomb{\textsf{Y}}

\newcommand{\tree}[2][]{\pstree[#1]{\TR{#2}}}
\newcommand{\pssetcomptree}{\psset{levelsep=4ex,linewidth=0.5pt}}

\newcommand{\arclabel}[1]{\mput*{\mbox{{\small $#1$}}}}

\theoremstyle{plain}
\newtheorem{theorem}[thm]{Theorem}
\newtheorem{corollary}[thm]{Corollary}
\newtheorem{lemma}[thm]{Lemma}
\newtheorem{proposition}[thm]{Proposition}

\theoremstyle{definition}
\newtheorem{definition}[thm]{Definition}

\newtheorem{example}[thm]{Example}

\theoremstyle{remark}
\newtheorem{remark}[thm]{Remark}

\newcommand\stentail{\vdash_{\sf st}} 
\newcommand\sentail{\vdash_{\sf s}} 
\newcommand\asappentail{\vdash_{\sf asa}} 
\newcommand\lsentail{\vdash_{\sf l}} 

\newcommand\enable\vdash 

\newcommand\Nodes{N}
\newcommand\INodes{IN}
\newcommand\LNodes{L}

\def\doi{5 (1:3) 2009}
\lmcsheading%
{\doi}
{1--38}
{}
{}
{Apr.~22, 2008}
{Feb.~19, 2009}
{}   

\begin{document}

\title[The safe lambda calculus]{The safe lambda calculus}

\author[W.~Blum]{William Blum}   
\address{Oxford University Computing Laboratory --
School of Informatics, University of Edinburgh, UK}    
\email{william.blum@comlab.ox.ac.uk}  

\author[C.-H.~L.~Ong]{C.-H.~Luke~Ong}   
\address{Oxford University Computing Laboratory, Oxford, UK} 
\email{luke.ong@comlab.ox.ac.uk}  




\titlecomment{Some of the results presented here were first published in TLCA proceedings \cite{blumong:safelambdacalculus}}


\begin{abstract}
  Safety is a syntactic condition of higher-order grammars that
  constrains occurrences of variables in the production rules
  according to their type-theoretic order. In this paper, we introduce
  the \emph{safe lambda calculus}, which is obtained by transposing
  (and generalizing) the safety condition to the setting of the
  simply-typed lambda calculus. In contrast to the original definition
  of safety, our calculus does not constrain types (to be
  homogeneous). We show that in the safe lambda calculus, there is no
  need to rename bound variables when performing substitution, as
  variable capture is guaranteed not to happen.  We also propose an
  adequate notion of $\beta$-reduction that preserves safety.  In the
  same vein as Schwichtenberg's 1976 characterization of the
  simply-typed lambda calculus, we show that the numeric functions
  representable in the safe lambda calculus are exactly the
  multivariate polynomials; thus conditional is not definable. We
  also give a characterization of representable word functions.
  We then study the complexity of deciding beta-eta equality of two safe simply-typed terms and show that this
  problem is PSPACE-hard.
  Finally we give a game-semantic analysis of safety: We show that
  safe terms are denoted by \emph{P-incrementally justified
    strategies}. Consequently pointers in the game semantics of safe
  $\lambda$-terms are only necessary from order 4 onwards.
\end{abstract}

    \maketitle              

    \section*{Introduction}

\subsection*{Background}

The \emph{safety condition} was introduced by Knapik, Niwi{\'n}ski and
Urzyczyn at FoSSaCS 2002 \cite{KNU02} in a seminal study of the
algorithmics of infinite trees generated by higher-order grammars. The
idea, however, goes back some twenty years to Damm \cite{Dam82} who
introduced an essentially equivalent\footnote{See de Miranda's
 thesis \cite{demirandathesis} for a proof.} syntactic
restriction (for generators of word languages) in the form of
\emph{derived types}.
A higher-order grammar (that is assumed to be \emph{homogeneously
  typed}) is said to be \emph{safe} if it obeys certain syntactic
conditions that constrain the occurrences of variables in the
production (or rewrite) rules according to their type-theoretic
order. Though the formal definition of safety is somewhat intricate,
the condition itself is manifestly important. As we survey in the
following, higher-order \emph{safe} grammars capture fundamental
structures in computation and offer clear algorithmic advantages:

\begin{enumerate}[$\bullet$]
\item \emph{Word languages}. Damm and Goerdt \cite{DG86} have shown
  that the word languages generated by order-$n$ \emph{safe} grammars
  form an infinite hierarchy as $n$ varies over the natural numbers.
  The hierarchy gives an attractive classification of the
  semi-decidable languages: Levels 0, 1 and 2 of the hierarchy are
  respectively the regular, context-free, and indexed languages (in
  the sense of Aho \cite{Aho68}), although little is known about
  higher orders.

  Remarkably, for generating word languages, order-$n$ \emph{safe}
  grammars are equivalent to order-$n$ pushdown automata \cite{DG86},
  which are in turn equivalent to order-$n$ indexed grammars
  \cite{Mas74,Mas76}.

\item \emph{Trees}. Knapik \emph{et al.} have shown that the Monadic
  Second Order (MSO) theories of trees generated by \emph{safe}
  (deterministic) grammars of every finite order are
  decidable\footnote{It has recently been shown
    \cite{OngLics2006} that trees generated by \emph{unsafe}
    deterministic grammars (of every finite order) also have decidable
    MSO theories. More precisely, the MSO theory of trees generated by order-$n$
recursion schemes is $n$-EXPTIME complete.}.

  They have also generalized the equi-expressivity result due to Damm
  and Goerdt \cite{DG86} to an equivalence result with respect to
  generating trees: A ranked tree is generated by an order-$n$ \emph{safe}
  grammar if and only if it is generated by an order-$n$ pushdown
  automaton.

\item \emph{Graphs}. Caucal \cite{Cau02} has shown that the MSO
  theories of graphs generated\footnote{These are precisely the
    configuration graphs of higher-order pushdown systems.} by
  \emph{safe} grammars of every finite order are decidable. Recently
  Hague \emph{et al.} have
  shown that the MSO theories of graphs generated by order-$n$
  \emph{unsafe} grammars are undecidable, but deciding their modal
  mu-calculus theories is $n$-EXPTIME complete \cite{hmos-lics08}.
\end{enumerate}

\subsection*{Overview}

In this paper, we examine the safety condition in the
setting of the lambda calculus. Our first task is to transpose it to
the lambda calculus and express it as an appropriate sub-system of
the simply-typed theory. A first version of the \emph{safe lambda
  calculus} has appeared in an unpublished technical report
\cite{safety-mirlong2004}. Here we propose a more general and cleaner
version where terms are no longer required to be homogeneously typed
(see Section~\ref{sec:safe} for a definition). The formation rules of
the calculus are designed to maintain a simple invariant: Variables
that occur free in a safe $\lambda$-term have orders no smaller than
that of the term itself.  We can now explain the sense in which the
safe lambda calculus is safe by establishing its salient property: No
variable capture can ever occur when substituting a safe term into
another. In other words, in the safe lambda calculus, it is
\emph{safe} to use capture-\emph{permitting} substitution when
performing $\beta$-reduction.

There is no need for new names when computing $\beta$-reductions of
safe $\lambda$-terms, because one can safely ``reuse'' variable names
in the input term. Safe lambda calculus is thus cheaper to compute in
this na\"ive sense. Intuitively one would expect the safety constraint
to lower the expressivity of the simply-typed lambda calculus. Our
next contribution is to give a precise measure of the expressivity
deficit of the safe lambda calculus. An old result of Schwichtenberg
\cite{citeulike:622637} says that the numeric functions representable
in the simply-typed lambda calculus are exactly the multivariate
polynomials \emph{extended with the conditional function}.  In the
same vein, we show that the numeric functions representable in the
safe lambda calculus are exactly the multivariate polynomials.

Our last contribution is to give a game-semantic account of the safe
lambda calculus.
Using a correspondence result relating the game semantics of a
$\lambda$-term $M$ to a set of \emph{traversals} \cite{OngLics2006}
over a certain abstract syntax tree of the $\eta$-long form of $M$
(called \emph{computation tree}), we show that safe terms are denoted
by \emph{P-incrementally justified strategies}. In such a strategy,
pointers emanating from the P-moves of a play are uniquely
reconstructible from the underlying sequence of moves and the pointers
associated to the O-moves therein: Specifically, a P-question always
points to the last pending O-question (in the P-view) of a greater
order. Consequently pointers in the game semantics of safe
$\lambda$-terms are only necessary from order 4 onwards. Finally we
prove that a $\beta$-normal $\lambda$-term is \emph{safe}
if and only if its strategy denotation is (innocent and)
\emph{P-incrementally justified}.

\section{The safe lambda calculus}
\label{sec:safe}
\subsection*{Higher-order safe grammars}
We first present the safety restriction as it was originally defined
\cite{KNU02}. We consider simple types generated by the grammar $A \,
::= \, o \; | \; A \typear A$. By convention, $\rightarrow$ associates
to the right. Thus every type can be written as $A_1 \typear \cdots
\typear A_n \typear o$, which we shall abbreviate to $(A_1, \cdots,
A_n, o)$ (in case $n = 0$, we identify $(o)$ with $o$).
We will also use the notation $A^n \typear B$ for every types $A, B$ and positive natural number $n>0$ defined by induction as: $A^1 \typear B = A \typear B$  and $A^{n+1} \typear B = A\typear (A^n \typear B)$. The \emph{order} of a type is given by $\ord{o} = 0$ and $\ord(A \typear
B) = \max(\ord{A}+1, \ord{B})$. We assume an infinite set of typed
variables. The order of a typed term or symbol is defined to be the
order of its type. The set of \emph{applicative terms} over a set of typed symbols is defined as its closure under the application operation (\ie, if $M : A\rightarrow B$ and $N :A$ are in the closure then so does $M N :B$).

A (higher-order) \defname{grammar} is a tuple $\langle
\Sigma, \mathcal{N}, \mathcal{R}, S \rangle$, where $\Sigma$ is a
ranked alphabet (in the sense that each symbol $f \in \Sigma$ is assumed to have type $o^r\typear o$ where $r$ is the \emph{arity} of $f$) of \emph{terminals}; $\mathcal{N}$ is a finite set of typed
\emph{non-terminals}; $S$ is a distinguished ground-type symbol of
$\mathcal{N}$, called the start symbol; $\mathcal{R}$ is a finite set
of production (or rewrite) rules, one for each non-terminal $F : (A_1,
\ldots, A_n, o) \in \mathcal{N}$, of the form $ F z_1 \ldots z_m
\rightarrow e$ where each $z_i$ (called \emph{parameter}) is a
variable of type $A_i$ and $e$ is an applicative term of type $o$
generated from the typed symbols in $\Sigma \union \mathcal{N} \union \{z_1,
\ldots, z_m \}$. We say that the grammar is \emph{order-$n$} just in
case the order of the highest-order non-terminal is $n$.

We call \defname{higher-order recursion scheme} a higher-order
grammar that is deterministic (\ie, for each non-terminal $F \in
\mathcal{N}$ there is exactly one production rule with $F$ on the
left hand side). Higher-order recursion schemes are used as
generators of infinite trees. The \defname{tree generated by a
recursion scheme} $G$ is a possibly infinite applicative term, but
viewed as a $\Sigma$-labelled tree; it is \emph{constructed from the
terminals in $\Sigma$}, and is obtained by unfolding the rewrite
rules of $G$ \emph{ad infinitum}, replacing formal by actual
parameters each time, starting from the start symbol $S$. See
e.g.~\cite{KNU02} for a formal definition.

\pssetcomptree
\parpic[r]{
\raisebox{-15pt}
{$\tree[levelsep=3ex,nodesep=1pt,treesep=1cm,linewidth=0.5pt]{g}
{  \TR{a}
    \tree{g}{\TR{a} \tree{h}{\tree{h}{\vdots}}}
}$}
}
\begin{example}\rm\label{eg:running}
  Let $G$ be the following order-2 recursion scheme:
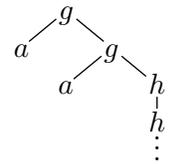
\[\begin{array}{rll}
  S & \rightarrow & H \, a\\
  H \, z^o & \rightarrow & F \, (g \,
  z)\\
  F \, \phi^{(o, o)} & \rightarrow & \phi \, (\phi \, (F \, h))\\
\end{array}\]
where the arities of the terminals $g, h, a$ are $2, 1, 0$ respectively.
The tree generated by $G$ is defined by the infinite term $g \, a \, (g \, a \, (h \, (h \, (h \,
\cdots))))$.

\end{example}

A type $(A_1, \cdots, A_n, o)$ is said to be \defname{homogeneous} if
$\ord{A_1} \geq \ord{A_2}\geq \cdots \geq \ord{A_n}$, and each $A_1$,
\ldots, $A_n$ is homogeneous \cite{KNU02}.  We reproduce the following
Knapik et al.'s definition \cite{KNU02}.

\begin{definition}[Safe grammar]\rm
\label{def:safegrammar}
  (All types are assumed to be homogeneous.) A term of order $k > 0$
  is \emph{unsafe} if it contains an occurrence of a parameter of
  order strictly less than $k$, otherwise the term is \emph{safe}. An
  occurrence of an unsafe term $t$ as a subexpression of a term $t'$
  is \emph{safe} if it is in the context $\cdots (ts) \cdots$,
  otherwise the occurrence is \emph{unsafe}. A grammar is
  \defname{safe} if no unsafe term has an unsafe occurrence at a
  right-hand side of any production.
\end{definition}

\begin{example}\begin{inparaenum}[(i)] \item Take $H : ((o, o), o)$ and $f : (o, o, o)$; the
    following rewrite rules are unsafe (In each case we underline the
    unsafe subterm that occurs unsafely):
\[\begin{array}{rll}
G^{(o, o)} \, x & \quad \rightarrow \quad & H \, \underline{(f \, {x})} \\
F^{((o, o), o, o, o)} \, z \, x \, y & \quad \rightarrow \quad & f \, (F \, \underline{(F \, z
\, {y})} \, y \, (z \, x) ) \, x
\end{array}\]
\item The order-2 grammar defined in Example~\ref{eg:running} is
  unsafe.
\end{inparaenum}
\end{example}

\subsection*{Safety adapted to the lambda calculus}
We assume a set $\Xi$ of higher-order constants. We use sequents of
the form $\Gamma \vdash_\$^\Xi M : A$ to represent term-in-context
where $\Gamma$ is the context and $A$ is the type of $M$. For
convenience, we shall omit the superscript from $\sentail^\Xi$
whenever the set of constants $\Xi$ is clear from the context. The
subscript in $\vdash_\$^\Xi$ specifies which type system is used to form the judgement: We use the subscript `st' to refer to
the traditional system of rules of the Church-style simply-typed
lambda calculus augmented with constants from $\Xi$. We will introduce a new subscripts for each type system that we define. For simplicity we write $(A_1, \cdots, A_n, B)$ to mean
$A_1 \typear \cdots \typear A_n \typear B$, where $B$ is not
necessarily ground.

\begin{definition}
\label{def:safelambda}
\begin{inparaenum}[(i)]
\item The \defname{safe lambda calculus} is a sub-system of the
  simply-typed lambda calculus. It is defined as the set of judgements of the form $\Gamma \sentail M : A$ that are derivable from the following Church-style system of rules:
$$ \rulename{var} \ \rulef{}{x : A\sentail x : A} \qquad
\rulename{const} \ \rulef{}{\sentail f : A}~f \in \Xi \qquad
\rulename{wk} \ \rulef{\Gamma \sentail M : A}{\Delta \sentail M : A} \quad
\Gamma \subset \Delta$$

$$ \rulename{app_{as}} \ \rulef{\Gamma \asappentail M : A\typear B
\quad \Gamma \sentail N : A} {\Gamma \asappentail M\, N : B}
\qquad
\rulename{\delta} \ \rulef{\Gamma \sentail M : A}{\Gamma \asappentail M : A}
$$

$$ \rulename{app} \ \rulef{\Gamma \asappentail M : A\typear B
\quad \Gamma \sentail N : A} {\Gamma \sentail M\, N : B} \quad \ord{B} \leq
\ord{\Gamma}$$

$$ \rulename{abs} \ \rulef{\Gamma, x_1 : A_1, \ldots, x_n : A_n
  \asappentail M : B} {\Gamma \sentail \lambda x_1^{A_1} \ldots x_n^{A_n} . M :
  (A_1, \ldots ,A_n,B)} \quad \ord(A_1, \ldots ,A_n,B) \leq
\ord{\Gamma}$$
\smallskip

\noindent where $\ord{\Gamma}$ denotes the set $\{ \ord{y} : y \in
\Gamma \}$ and ``$c \leq S$'' means that $c$ is a lower-bound of the
set $S$. The subscripts in $\sentail$ and $\asappentail$ stand
for ``safe'' and ``almost safe application''.

\item The sub-system that is defined by the same rules in
(i), such that all types that occur in them are homogeneous, is called
the \defname{homogeneous safe lambda calculus}.

\item We say that a \emph{term} $M$ is \defname{safe} if the judgement $\Gamma \sentail M : T$ is derivable in the safe lambda calculus for some context $\Gamma$ and type $T$.
\end{inparaenum}
\end{definition}

The safe lambda calculus deviates from the standard definition of
the simply-typed lambda calculus in a number of ways. 
First the rule $\rulename{abs}$ can abstract several variables at once. (Of course this feature alone does not alter expressivity.) Crucially,
the side conditions in the application rule and abstraction rule
require the variables in the typing context to have orders no
smaller than that of the term being formed.  We do not impose any
constraint on types. In particular, type-homogeneity, which was an
assumption of the original definition of safe grammars \cite{KNU02},
is not required here. Another difference is that we allow
$\Xi$-constants to have
arbitrary higher-order types.
\bigskip

\begin{example}[Kierstead terms]
\label{ex:kierstead}
Consider the terms $M_1 = \lambda f^{((o,o),o)} . f (\lambda x^o . f (\lambda y^o . y
))$ and $M_2 = \lambda f^{((o,o),o)} . f (\lambda x^o . f (\lambda y^o .x ))$. The term $M_2$ is not safe because in the subterm $f (\lambda y^o . x)$, the free variable $x$ has order $0$ which
is smaller than $\ord(\lambda y^o . x) = 1$.  On the other hand, $M_1$
is safe.
\end{example}

It is easy to see that valid typing judgements of the safe lambda
calculus satisfy the following simple invariant:
\begin{lemma}
\label{lem:ordfreevar}
If $\Gamma \sentail M : A$ then every variable in $\Gamma$ occurring
free in $M$ has order at least $\ord M$.
\end{lemma}

\begin{definition}
A term is an \defname{almost safe applications}
if it is safe or if it is of the form $N_1 \ldots N_m$ for some $m\geq 1$ where $N_1$ is not an application and for every $1 \leq i\leq m$, $N_i$ is safe.

A term is \defname{almost safe} if either it is an almost safe application, or if it is of the form
$\lambda x_1^{A_1} \ldots x_n^{A_n}. M$ for $n\geq 1$ and some almost safe application $M$.
\end{definition}
An \emph{almost safe application} is not necessarily safe but it can be used to form a safe term by applying sufficiently many safe terms to it. An \emph{almost safe term} can be turned into a safe term by either applying sufficiently many safe terms (if it is an application), or
by abstracting sufficiently many variables (if it is an abstraction).

We have the following immediate lemma:
\begin{lemma}
\label{lem:almostsafeapp_is_appplicative_safe}
A term $M$ is
\begin{enumerate}[\em(i)]
\item an \emph{almost safe application} iff there is a derivation of $\Gamma \asappentail M : T$ for some $\Gamma, T$;

\item \emph{almost safe} iff $\Gamma \asappentail M : T$ or if $M\equiv \lambda x_1^{A_1} \ldots x_n^{A_n}. N$ and $\Gamma \asappentail N : T$ for some $\Gamma, T$.
\end{enumerate}
\end{lemma}
In particular, terms constructed with the rule \rulenamet{app_{as}} are almost safe applications.
\bigskip

When restricted to the homogeneously-typed sub-system, the safe
lambda calculus captures the original notion of safety due to Knapik
\emph{et al.}~in the context of higher-order grammars:

\begin{proposition}
\label{prop:safegram_safelmd}
 Let $G = \langle \Sigma, \mathcal{N}, \mathcal{R},
  S \rangle$ be a grammar and let $e$ be an applicative term generated
  from the symbols in $\mathcal{N} \cup \Sigma \cup \makeset{z_1^{A_1},
    \cdots, z_m^{A_m}}$.  A rule $F z_1 \ldots z_m \rightarrow e$ in
  $\mathcal{R}$ is safe (in the original sense of Knapik \emph{et al.}) if and only if $ z_1 : A_1, \cdots, z_m : A_m
  \sentail^{\Sigma \cup \mathcal{N}} e : o$ is a valid typing judgement
  of the \emph{homogeneous} safe lambda calculus.
\end{proposition}
\begin{proof}
We show by induction that \begin{asparaenum}[(i)]
\item  $z_1,\ldots, z_m \asappentail t:A$ is a valid judgement of the homogeneous safe lambda calculus containing no abstraction if and only if in the Knapik sense, all the occurrences of unsafe subterms of $t$ are safe occurrences.
\item $z_1,\ldots, z_m \sentail t:A$ is a valid judgement of the homogeneous safe lambda calculus containing no abstraction if and only if in the Knapik sense, all the occurrences of unsafe subterms of $t$ are safe occurrences, and all parameters occurring in $t$ have order greater than $\ord{t}$. \end{asparaenum}
The constant and variable rule are trivial. Application case: By definition, a term $t_0 \ldots t_n$ is Knapik-safe iff for all $0\leq i \leq n$, all the occurrences of unsafe subterms of $t_i$ are safe occurrences (in the Knapik sense), and for all $1\leq j \leq n$, the operands occurring in $t_j$ have order greater than $\ord{t_j}$. The \rulenamet{app_{as}} rule and the induction hypothesis permit us to conclude.

Now since $e$ is an applicative term of \emph{ground type}, the previous result gives:
$z_1,\ldots, z_m \sentail e:o$ is a valid judgement of the homogeneous safe lambda calculus iff all the occurrences of unsafe subterms of $e$ are safe occurrences, which by definition of Knapik-safety is in turn equivalent to saying that the rule $F z_1 \ldots z_m \rightarrow e$ is safe.
\end{proof}

\emph{In what sense is the safe lambda calculus safe?} It is an
elementary fact that when performing $\beta$-reduction in the lambda
calculus, one must use capture-\emph{avoiding} substitution, which
is standardly implemented by renaming bound variables afresh upon
each substitution. In the safe lambda calculus, however, variable
capture can never happen (as the following lemma shows).
Substitution can therefore be implemented simply by
capture-\emph{permitting} replacement, without any need for variable
renaming. In the following, we write $M\captsubst{N}{x}$ to denote
the capture-\emph{permitting} substitution\footnote{This
substitution is done by textually replacing all free occurrences of
$x$ in $M$ by $N$ without performing variable renaming.  In
particular for the abstraction
  case we have
$(\lambda y_1\ldots y_n . M)\captsubst{N}{x} = \lambda y_1\ldots y_n . M\captsubst{N}{x}$ when $x\not\in
  \{ y_1\ldots y_n \}$.}
of $N$ for $x$ in $M$.

\begin{lemma}[No variable capture]\label{lem:nvc}
\label{lem:nocapture} There is no variable capture when performing
capture-permitting substitution of $N$ for $x$ in $M$ provided that
$\Gamma, x:B \sentail M : A$ and $\Gamma \sentail  N : B$ are valid
judgements of the safe lambda calculus.
\end{lemma}

\begin{proof}
  We proceed by structural induction on $M$. The variable, constant and
  application cases are trivial. For the abstraction case, suppose $M \equiv \lambda \overline{y}. R$ where $\overline{y} = y_1 \ldots y_p$. If $x \in \overline{y}$ then $M \captsubst{N}{x} = M$ and there is no variable capture.

 Otherwise, $x \not\in \overline{y}$. By Lemma \ref{lem:almostsafeapp_is_appplicative_safe} $R$ is of the
  form $M_1 \ldots M_m$ for some $m\geq 1$ where $M_1$ is not an application and for every $1 \leq i\leq m$, $M_i$ is safe.
 Thus we have $M \captsubst{N}{x} \equiv \lambda \overline{y} . M_1 \captsubst{N}{x} \ldots M_m \captsubst{N}{x}$.  Let $i\in\{1..m\}$. By the induction hypothesis there is no variable capture in $M_i \captsubst{N}{x}$.  Thus variable capture can only happen if the following two conditions are met: (i) $x$ occurs freely in $M_i$, (ii) some variable $y_i$ for $1 \leq i \leq p$ occurs freely in $N$. By Lemma \ref{lem:ordfreevar}, (ii) implies $\ord{y_i} \geq \ord{N} = \ord{x}$ and since $x \not \in \overline{y}$, condition (i) implies that $x$ occurs freely in the safe term $\lambda \overline{y}. R$
  thus by Lemma \ref{lem:ordfreevar} we have $ \ord{x} \geq
  \ord{\lambda \overline{y} . R} \geq 1+ \ord{y_i} > \ord{y_i}$ which  gives a contradiction.
\end{proof}

\begin{remark}
  A version of the No-variable-capture Lemma also holds in safe
  grammars, as is implicit in (for example Lemma 3.2 of) the original
  paper \cite{KNU02}.
\end{remark}

\begin{example}
  In order to contract the $\beta$-redex in the term
\[f:(o,o,o),x:o
  \stentail (\lambda \varphi^{(o,o)} x^o . \varphi \, x) (\underline{f \,
    x}) : (o,o)\] one should rename the bound variable $x$ to a fresh name to
  prevent the capture of the free occurrence of $x$ in the underlined term during substitution. Consequently, by the previous lemma,
  the term is not safe (because $\ord{x} = 0 < 1
  = \ord{f x}$).
\end{example}

Note that $\lambda$-terms that `satisfy' the No-variable-capture
Lemma are not necessarily safe. For instance the $\beta$-redex in
$\lambda y^o z^o. (\lambda x^o .y) z$ can be contracted using
capture-permitting substitution, even though the term is not safe.
\bigskip

\emph{Related work:} In her thesis \cite{demirandathesis}, de Miranda proposed a different notion of safe lambda calculus. This notion corresponds to (a less general version of) our notion of \emph{homogeneous} safe lambda calculus. It can be showed that for pure applicative terms (\ie, with no lambda-abstraction) the two systems coincide. In particular a version of Proposition \ref{prop:safegram_safelmd} also holds in de Miranda's setting \cite{demirandathesis}. In the presence of lambda abstraction, however, our system is less restrictive. For instance the term $\lambda f^{(o,o,o)} x^o.  f x : (o,o)$ is typable in the homogeneous safe lambda calculus but not in the safe lambda calculus \emph{\`a la} de Miranda. One can show that de Miranda's system is in fact equivalent to the \emph{homogeneous long-safe lambda calculus} (\ie, the restriction of the system of Def.\ \ref{dfn:longsafe} to homogeneous types).

\subsection*{Safe beta reduction}

From now on we will use the standard notation $M\subst{N}{x}$ to
denote the substitution of $N$ for $x$ in $M$.  It is understood that,
provided that $M$ and $N$ are safe, this substitution is
capture-permitting.

\begin{lemma}[Substitution preserves safety]
\label{lem:subst_preserves_safety}
Let $\Gamma \sentail N : B$. Then
\begin{enumerate}[\em(i)]
  \item $\Gamma, x :B \sentail M : A$ implies $\Gamma \sentail M[N/x] : A$;
  \item $\Gamma, x :B \asappentail M : A$ implies $\Gamma \asappentail M[N/x] : A$.
\end{enumerate}
\end{lemma}
This is proved by an easy induction on the structure of the safe term $M$.
\smallskip

It is desirable to have an appropriate notion of reduction for our
calculus. However the standard $\beta$-reduction rule is not
adequate. Indeed, safety is not preserved by $\beta$-reduction as
the following example shows. Suppose that $w,z : o$ and $f :
(o,o,o) \in \Sigma$ then the safe term $(\lambda x^o y^o . f x y) z w$
$\beta$-reduces to $(\underline{\lambda y^o . f z y}) w$, which is
unsafe since the underlined first-order subterm contains a free
occurrence of the ground-type variable $z$. However if we perform one more
reduction we obtain the safe term $f z w$. This suggests
simultaneous contraction of ``consecutive'' $\beta$-redexes. In
order to define this notion of reduction we first introduce the
corresponding notion of redex.

In the simply-typed lambda calculus a redex is a term of the form
$(\lambda x . M) N$. In the safe lambda calculus, a redex is a
succession of several standard redexes:

\begin{definition}\rm
A \defname{safe redex} is an almost safe application of the form
$$(\lambda x_1^{A_1} \ldots x_n^{A_n}. M) N_1 \ldots N_l$$ for $l,n\geq 1$ such that $M$ is an almost safe application.
(Consequently each $N_i$ is safe as well as $\lambda x_1^{A_1} \ldots x_n^{A_n} . M$, and $M$ is either safe or is an application of safe terms.)
\end{definition}
For instance, in the case $n<l$, a safe redex has a derivation tree of the following form:
\def\defaultHypSeparation{}
\begin{prooftree}
    \AxiomC{\ldots}
  \UnaryInfC{$\Gamma', \overline{x}:\overline{A} \sentail M : (A_{n+1}, \ldots, A_l, B)$}
  \RightLabel{\rulenamet{abs}}
  \UnaryInfC{$\Gamma' \sentail \lambda x_1^{A_1} \ldots x_n^{A_n} . M : (A_1, \ldots, A_l, B)$}
  \RightLabel{\rulenamet{wk}}
  \UnaryInfC{$\Gamma \sentail \lambda x_1^{A_1} \ldots x_n^{A_n} . M : (A_1, \ldots, A_l, B)$}
  \RightLabel{\rulenamet{\delta}}
  \UnaryInfC{$\Gamma \asappentail \lambda x_1^{A_1} \ldots x_n^{A_n} . M : (A_1, \ldots, A_l, B)$}
  \AxiomC{\ldots}
  \UnaryInfC{$\Gamma \sentail N_1 :A_1$}
  \RightLabel{\rulenamet{app_{as}}}
   \BinaryInfC{$\Gamma \asappentail (\lambda x_1^{A_1} \ldots x_n^{A_n} . M) N_1 : (A_2, \ldots A_l, B)$}
  \noLine\UnaryInfC{\vdots\raisebox{0.5cm}{}}
  \RightLabel{\rulenamet{app_{as}}}
  \UnaryInfC{$\Gamma \asappentail (\lambda x_1^{A_1} \ldots x_n^{A_n} . M) N_1 \ldots N_{l-1} : (A_l, B)$}
  \AxiomC{\ldots}
  \UnaryInfC{$\Gamma \sentail N_l :A_l$}
  \RightLabel{\rulenamet{app}}
  \BinaryInfC{$\Gamma \sentail (\lambda x_1^{A_1} \ldots x_n^{A_n} . M) N_1 \ldots N_l : B$}
\end{prooftree}
\smallskip

A \emph{safe redex} is by definition an almost term,
but it is not necessarily a \emph{safe term}. For instance the term $(\lambda x^o y^o. x) z$ is a safe redex but it is only an \emph{almost} safe term.
The reason why we call such redexes ``safe'' is because when they occur within a safe term, it is possible to contract them without braking the safety of the whole term. Before showing this result, we first need to define how to contract safe redexes:
\begin{definition}[Redex contraction]\rm
\label{dfn:saferedex_contraction} We use the abbreviations $\overline{x} =
x_1 \ldots x_n$, $\overline{N} = N_1 \ldots N_l$. The relation
$\beta_s$ (when viewed as a function) is defined on the set of \emph{safe redexes} as follows:
\begin{eqnarray*}
  \beta_s &=&
  \{  \ (\lambda x_1^{A_1} \ldots x_n^{A_n} . M) N_1 \ldots N_l \mapsto \lambda x_{l+1}^{A_{l+1}} \ldots x_n^{A_{n}}. M\subst{\overline{N}}{x_1 \ldots x_l} \ | \ n> l  \} \\
  &\cup&
  \{ \ (\lambda x_1^{A_1} \ldots x_n^{A_n} . M) N_1 \ldots N_l \mapsto M\subst{N_1 \ldots N_n}{\overline{x}} N_{n+1} \ldots N_l
  \ | \ n\leq l \} \ .
\end{eqnarray*}
where $M\subst{R_1 \ldots R_k}{z_1 \ldots z_k}$ denotes the simultaneous substitution in $M$ of $R_1$,\ldots,$R_k$ for $z_1, \ldots, z_k$.
\end{definition}

\begin{lemma}[$\beta_s$-reduction preserves safety]
\label{lem:betas_preserves_safety}
Suppose that $M_1\, \beta_s\, M_2$. Then
\begin{enumerate}[\em(i)]
  \item $M_2$ is almost safe;
  \item If $M_1$ is safe then so is $M_2$.
\end{enumerate}
\end{lemma}
\begin{proof}
Let $M_1\, \beta_s\, M_2$ for some safe redex $M_1$ and term $M_2$ of type $A$. By definition, $M_1$ is of the form $(\lambda x_1^{B_1} \ldots x_n^{B_n} . M) N_1 \ldots N_l $ for some safe terms $N_1$, \ldots, $N_l$ and almost safe term $M$ of type $C$ such that $(\lambda x_1^{B_1} \ldots x_n^{B_n} . M)$ is safe.
\begin{enumerate}[$-$]
\item
Suppose $n>l$ then $A = (B_{l+1}, \ldots, B_n, C)$.
(i) By the Substitution Lemma
\ref{lem:subst_preserves_safety}, the term $M\subst{\overline{N}}{x_1
\ldots x_l}$ is an almost safe application: we have $\Gamma, x_{l+1}:B_{l+1}, \ldots x_n :B_{n}\asappentail M\subst{\overline{N}}{x_1
\ldots x_l} : C$. (Indeed, if $M$ is safe then we apply the Substitution Lemma once; otherwise it is of the form $R_1 \ldots R_q$ where $R_i$ is a safe term and we apply the lemma on each $R_i$.)
Thus by definition, $\lambda x_{l+1}^{B_{l+1}} \ldots x_n^{B_n} . M\subst{\overline{N}}{x_1
\ldots x_l} \equiv M_2$ is almost safe.

(ii) Suppose that $M_1$ is safe. W.l.o.g.~we can assume that the last rule used to form  $M_1$ is \rulenamet{app} (and not the weakening rule
\rulenamet{wk}), thus the variables of the typing context $\Gamma$ are precisely the free variables of $M_1$, and Lemma \ref{lem:ordfreevar} gives us $\ord{A} \leq \ord{\Gamma}$. This allows us to use the rule \rulenamet{abs} to form the safe term-in-context $\Gamma
\sentail \lambda x_{l+1}^{B_{l+1}} \ldots x_n^{B_n} . M\subst{\overline{N}}{x_1
\ldots x_l} \equiv M_2 : A$.

\item Suppose $n \leq l$. (i) Again by the Substitution Lemma
we have that $M\subst{N_1 \ldots N_n}{\overline{x}}$ is an almost safe application: $\Gamma \asappentail M\subst{N_1 \ldots N_n}{\overline{x}} : C$. If $n=l$ then the proof is finished; otherwise ($n<l$) we further apply the rule \rulenamet{app_{as}} $l-n$ times which gives us the almost safe application $\Gamma \asappentail M_2 :A$.

(ii) Suppose that $M_1$ is safe.  If $n=l$ then $M_2\equiv M\subst{N_1 \ldots N_n}{\overline{x}}$ is safe by the Substitution Lemma;
If $n<l$ then we obtain the judgement $\Gamma \sentail M_2 :A$ by
applying the rule \rulenamet{app_{as}} $l-n-1$ times on $\Gamma \sentail M\subst{N_1 \ldots N_n}{\overline{x}} : C$ followed by one application of \rulenamet{app}.
\qedhere
\end{enumerate}
\end{proof}

We can now define a notion of reduction for safe terms.
\begin{definition}\rm
The \defname{safe $\beta$-reduction}, written $\betasred$, is the
compatible closure of the relation $\beta_s$ with respect to the
formation rules of the safe lambda calculus (\ie, it is the smallest relation such that
if $M_1\, \beta_s\, M_2$ and $C[M]$ is a safe term for some context $C[-]$ formed with the rules of the simply-typed lambda calculus then $C[M_1]\betasred C[M_2]$).
\end{definition}

\begin{lemma}[$\beta_s$-reduction preserves safety]
\label{lem:safered_preserves_safety}
If $\Gamma \sentail M_1 :A$ and $M_1\betasred M_2$ then $\Gamma \sentail M_2 :A$.
\end{lemma}
\begin{proof}
Follows from Lemma \ref{lem:betas_preserves_safety} by an easy induction.
\end{proof}

\begin{lemma} The safe reduction relation $\betasred$:
\begin{enumerate}[\em(i)]
\item is a subset of the transitive closure of $\betared$ ($\betasred \subset \betaredtr$);
\item is strongly normalizing;
\item has the unique normal form property;
\item has the Church-Rosser property.
\end{enumerate}
\end{lemma}
\begin{proof}
(i) Immediate from the definition: Safe $\beta$-reduction is just a multi-step $\beta$-reduction.
(ii) This is because $\betasred \subset \betaredtr$ and, $\betared$ is
strongly normalizing in the simply-typed $\lambda$-calculus. (iii) It is easy to see that if a safe term has a beta-redex if and only if it has a safe beta-redex (because a beta-redex can always be ``widen'' into consecutive beta-redex of the shape of those in Def.~\ref{dfn:saferedex_contraction}). Therefore the set
of $\beta_s$-normal forms is equal to the set of $\beta_s$-normal
forms. The uniqueness of $\beta$-normal form then implies the uniqueness
of $\beta_s$-normal form. (iv) is a consequence of (i) and (ii).
\end{proof}

\subsection*{Eta-long expansion}

The $\eta$-long normal form (or simply $\eta$-long form) of a term
is obtained by hereditarily $\eta$-expanding the body of every
lambda abstraction as well as every subterm occurring in an
\emph{operand position} (\ie, occurring as the second argument of
some occurrence of the binary application operator). Formally the
\defname{$\eta$-long form}, written $\elnf{M}$, of a (type-annotated) term $M$ of type
$(A_1,\ldots,A_n,o)$ with $n \geq 0$ is defined by cases according to
the syntactic shape of $M$:
\begin{eqnarray*}
  \elnf{\lambda x^\tau . N } &\equiv& \lambda x^\tau . \elnf{N} \\
  \elnf{x N_1 \ldots N_m } &\equiv& \lambda \overline{\varphi}^{\overline{A}} . x \elnf{N_1}\ldots \elnf{N_m} \elnf{\varphi_1} \ldots \elnf{\varphi_n} \\
  \elnf{(\lambda x^\tau . N) N_1 \ldots N_p } &\equiv& \lambda \overline{\varphi}^{\overline{A}} . (\lambda x^\tau . \elnf{N}) \elnf{N_1} \ldots \elnf{N_p} \elnf{\varphi_1} \ldots \elnf{\varphi_n}
\end{eqnarray*}
where $m \geq 0$, $p\geq 1$, $x$ is either a variable or constant, $\overline{\varphi} = \varphi_1 \ldots \varphi_n$ and each $\varphi_i : A_i$ is a fresh variable.
The binder notation `$\lambda \overline{\varphi}^{\overline A}$' stands for
`$\lambda \varphi_1^{A_1} \ldots \varphi_n^{A_n}$' if $n\geq 1$, and for `$\lambda$' (called the \emph{dummy lambda}) in the case $n=0$. The base case of this inductive definition lies in the second clause for $m=n=0$: $\elnf{x} \equiv \lambda. x$.

\begin{remark}
  This transformation does not introduce
  new redexes therefore the $\eta$-long normal form of a $\beta$-normal
  term is also $\beta$-normal.
\end{remark}

Let us introduce a new typing system:
\begin{definition}
\label{dfn:longsafe}
We define the set of \defname{long-safe terms}
by induction over the following system of rules:
  $$ \rulename{var_l} \ \rulef{}{x : A\lsentail x : A} \qquad
\rulename{const_l} \ \rulef{}{\lsentail f : A}\quad f \in \Xi \qquad
\rulename{wk_l} \ \rulef{\Gamma \lsentail M : A}{\Delta \lsentail M : A}\quad
\Gamma \subset \Delta$$

$$ \rulename{app_l} \ \rulef{\Gamma \lsentail M : (A_1,\ldots,A_n,B)
\quad
  \Gamma \lsentail N_1 : A_1 \quad \ldots \quad \Gamma \lsentail N_n : A_n
} {\Gamma \lsentail M N_1 \ldots N_n : B} \quad \ord B \leq
\ord \Gamma$$

$$ \rulename{abs_l} \ \rulef{\Gamma, x_1 : A_1, \ldots, x_n : A_n
  \lsentail M : B} {\Gamma \lsentail \lambda x_1^{A_1} \ldots x_n^{A_n} . M :
  (A_1, \ldots ,A_n,B)} \quad \ord(A_1, \ldots ,A_n,B) \leq
\ord \Gamma$$
\end{definition}
The subscript in $\lsentail$ stands for ``long-safe''. This terminology is deliberately suggestive of a forthcoming lemma. Note that long-safe terms are not necessarily in $\eta$-long normal form.

Observe that the system of rules from Def.~\ref{dfn:longsafe} is a sub-system of the typing system of Def.~\ref{def:safelambda} where the application rule is  restricted the same way as the abstraction rule (\ie, it can perform multiple applications at once provided that all the variables in the context of the resulting term have order greater than the order of the term itself). Thus we clearly have:
\begin{lemma}
\label{lem:longsafe_imp_safe}
If a term is long-safe then it is safe.
\end{lemma}
\smallskip

In general, long-safety is not preserved by $\eta$-expansion. For
instance we have
$\lsentail \lambda y^o z^o . y : (o,o,o)$ but
performing one eta-expansion produces the term $\lambda x^o . (\lambda y^o z^o . y) x : (o,o,o)$ which is not long-safe.
On the other hand, $\eta$-reduction (of one variable) preserves long-safety:

\begin{lemma}[$\eta$-reduction of one variable preserves long-safety]
\label{lem:etared_preserve_longsafety}
  $\Gamma \lsentail \lambda \varphi^\tau . M\, \varphi :A $ with $\varphi$ not
  occurring free in $s$ implies $\Gamma \lsentail M :A$.
\end{lemma}
\begin{proof}
  Suppose $\Gamma \lsentail \lambda \varphi^\tau . M\, \varphi :A$. If $M$ is an  abstraction then by construction of $M$ is necessarily safe.  If $M \equiv N_0 \ldots N_p$ with
  $p\geq 1$ then again, since $\lambda \varphi^\tau . N_0 \ldots N_p
  \varphi$ is safe, each of the $N_i$ is safe for $0 \leq i \leq p$
  and for every variable $z$ occurring free in $\lambda \varphi . M\, \varphi$, $\ord{z} \geq \ord(\lambda \varphi^\tau . M\, \varphi) = \ord M$. Since  $\varphi$ does not occur free in $M$, the terms $M$ and $\lambda \varphi^\tau . M\, \varphi$ have the same set of free variables, thus we can use the application rule to form $\Gamma' \lsentail N_0 \ldots N_p : A$ where $\Gamma'$ consists of the typing-assignments for the free variables of $M$. The weakening rules permits us to conclude $\Gamma \lsentail M :A$.
\end{proof}

\begin{lemma}[$\eta$-long expansion preserves long-safety]
\label{lem:longsafe_imp_elnf_longsafe}
$\Gamma \lsentail M :A$ then $\Gamma \lsentail \elnf{M} :A$.
\end{lemma}
\begin{proof}
 First we observe that for every variable or constant $x:A$ we have $x:A \lsentail \elnf{x} :A$. We show this by induction on $\ord{x}$.
It is verified for every ground type variable $x$
since $x = \elnf{x}$.
Step case: $x:A$ with $A=(A_1, \ldots, A_n,o)$ and $n>0$. Let $\varphi_i:A_i$ be fresh variables for $1\leq i\leq n$.
Since $\ord{A_i} < \ord{x}$ the induction hypothesis gives $\varphi_i :A_i \lsentail \elnf{\varphi_i} : A_i$. Using \rulenamet{wk_l} we obtain $x:A, \overline{\varphi} : \overline{A}
  \lsentail \elnf{\varphi_i} :A_i$.  The application rule gives $x :A, \overline{\varphi} : \overline{A} \lsentail x \elnf{\varphi_1} \ldots \elnf{\varphi_n}
  : o$ and the abstraction rule gives $ x :A \lsentail \lambda
  \overline{\varphi} . x \elnf{\varphi_1} \ldots \elnf{\varphi_n} =
  \elnf{x} :A$.

We now prove the lemma by induction on $M$.
The base case is covered by the previous observation.
\emph{Step case:}
\begin{enumerate}[$\bullet$]
\item $M \equiv x N_1 \ldots N_m$ with $x: (B_1, \ldots, B_m, A)$, $A = (A_1, \ldots, A_n, o)$ for some $m\geq 0$, $n>0$ and $N_i : B_i$ for $1 \leq i \leq
  m$.  Let $\varphi_i: A_i$ be fresh variables for $1\leq i \leq
  n$. By the previous observation we have $\varphi_i :A_i \lsentail \elnf{\varphi_i} :A_i$, the weakening rule then gives us $\Gamma , \overline{\varphi} : \overline{A}
  \lsentail \elnf{\varphi_i} : A_i$.  Since the judgement
  $\Gamma \lsentail x N_1 \ldots N_m : A$ is formed using the \rulenamet{app_l} rule, each $N_j$ must be long-safe for $1\leq j \leq m$, thus by the induction hypothesis we have $\Gamma \lsentail \elnf{N_j} : B_j$ and by weakening we get $\Gamma, \overline{\varphi} :\overline{A} \lsentail \elnf{N_j} : B_j$.  The \rulenamet{app_l}
  rule then gives $\Gamma, \overline{\varphi} :\overline{A} \lsentail x \elnf{N_1} \ldots \elnf{N_m} \elnf{\varphi_1} \ldots \elnf{\varphi_n} : o$. Finally
  the \rulenamet{abs_l} rule gives $\Gamma \lsentail \lambda \overline{\varphi} . x
  \elnf{N_1} \ldots \elnf{N_m} \elnf{\varphi_1} \ldots
  \elnf{\varphi_n} \equiv \elnf{M} : A$, the side-condition of \rulenamet{abs_l} being verified since $\ord{\elnf{s}} = \ord{s}$.

\item $M \equiv N_0 \ldots N_m$ where $N_0$ is an abstraction and $m\geq 1$.
The eta-long normal form is $\elnf{M} \equiv \lambda \overline{\varphi}. \elnf{N_0} \ldots \elnf{N_m} \elnf{\varphi_1}  \ldots \elnf{\varphi_n}$ for some fresh variables $\varphi_1$, \ldots, $\varphi_n$. Again, using the induction hypothesis we can easily derive $\Gamma \lsentail
 \elnf{M} : A$.

\item $M \equiv \lambda \overline{\eta}^{\overline{B}} . N $ where
$N$ of type $C$ and is not an abstraction. The induction hypothesis gives $\Gamma,
  \overline{\eta} : \overline{B} \lsentail \elnf{N} : C$ and using
\rulenamet{abs_l} we get $\Gamma \lsentail \lambda \overline{\eta} . \elnf{N} \equiv \elnf{M} : A$.  \qedhere
\end{enumerate}
\end{proof}
\smallskip

\begin{remark}\hfill
\begin{enumerate}[(i)]
\item
The converse of this lemma does not hold: performing
$\eta$-reduction over a large abstraction does not in general
preserve long-safety. (This does not contradict Lemma
  \ref{lem:etared_preserve_longsafety} which states that safety is
  preserved when performing $\eta$-reduction on an abstraction
  of a \emph{single} variable.) A counter-example is
 $\lambda f^{(o,o,o)} g^{((o,o,o),o)} . g(\lambda x^o . f
\underline{x})$, which is not long-safe but whose eta-normal form $\lambda f^{(o,o,o)}
g^{((o,o,o),o)} . g(\lambda x^o y^o. f x y)$ is long-safe.
There are also closed terms \emph{in eta-normal form}
that are not long-safe but have an $\eta$-long normal form that is long-safe!
Take for instance the closed $\beta\eta$-normal term  $\lambda
f^{(o,(o,o),o,o)} g^{((o,o),o,o,o),o)} . g(\lambda y^{(o,o)} x^o
. f \underline{x} y)$.

  \item After performing $\eta$-long expansion of a term, all the occurrences of the application rule are made long-safe. Thus if a term remains not long-safe after $\eta$-long expansion, this means that
  some variable occurrence is not bound by the
  first following application of the \rulenamet{abs} rule in the
  typing tree.
  \end{enumerate}
\end{remark}

\begin{lemma}
  \label{lem:safe_iff_etalong_lsafe}
  A simply-typed term is safe if and only if its $\eta$-long normal form is long-safe.
\end{lemma}
\begin{proof}
Let $\Gamma \stentail M : T$. We want to show that we have $\Gamma \sentail M : T$ if and only if $\Gamma \lsentail \elnf{M} : T$. The
`Only if' part can be proved by a trivial induction on the structure of
$\Gamma \sentail M : T$. For the `if' part we proceed by
induction on the structure of the simply-typed term $M$: The variable and constant cases are
trivial. Suppose that $M$ is an application of the form $x N_1
\ldots N_m : A$ for $m\geq 1$. Its $\eta$-long normal form is of the
form $x \elnf{N_1} \ldots \elnf{N_m} \elnf{\varphi_1} \ldots
\elnf{\varphi_m}: o$ for some fresh variables $\varphi_1$, \ldots
$\varphi_m$. By assumption this term is long-safe therefore we
have $\ord{A}\leq\ord{\Gamma}$ and for $1\leq i \leq m$,
$\elnf{N_i}$ is also long-safe. By the induction hypothesis this
implies that the $N_i$s are all safe. We can then form the judgement
$\Gamma \sentail x N_1 \ldots N_m : A$ using the rules
$\rulename{var}$ and $\rulename{\delta}$ followed by $m-1$
applications of the rule $\rulename{app_{as}}$ and one application
of $\rulename{app}$ (this is allowed since we have
$\ord{A}\leq\ord{\Gamma}$). The case $M\equiv (\lambda x. N) N_1
\ldots N_m$ for $m\geq 1$ is treated identically.

Suppose that $M \equiv \lambda \overline{x}^{\overline B} . N : A$. By assumption,
its  $\eta$-long n.f.\ $\lambda {\overline x}^{\overline B} {\overline \varphi}^{\overline C} .
\elnf{N} \elnf{\varphi_1} \ldots \elnf{\varphi_m}: A$ (for some
fresh variables $\overline\varphi = \varphi_1 \ldots \varphi_m$ and types
$\overline C = C_1 \ldots C_m$) is long-safe. Thus
we have $\ord{A}\leq\ord{\Gamma}$. Furthermore the long-safe subterm
$\elnf{N} \elnf{\varphi_1} \ldots \elnf{\varphi_m}$ is precisely the
eta-long normal form of $N\, \varphi_1 \ldots \varphi_m : o$ therefore by
the induction hypothesis we have that $N \varphi_1 \ldots \varphi_m
:o$ is safe. Since the $\varphi_i$'s are all safe (by rule
$\rulename{var}$), we can ``peel-off'' $m$ applications (performed using
the rules $\rulename{app_{as}}$ or $\rulename{app}$) from the sequent $\Gamma,
\overline{x}:\overline B,
\overline{\varphi}:\overline C \sentail N\, \varphi_1 \ldots
\varphi_m :o$ which gives us the sequent $\Gamma, \overline{x}:\overline B,
\overline{\varphi}:\overline C \asappentail N : A$. Since the
variables $\overline{\varphi}$ are fresh for $N$, we can further peel-off
applications of the weakening rule to obtain the judgement $\Gamma,
\overline{x}:\overline B \sentail N : A$.

Finally since we have $\ord{A}\leq\ord{\Gamma}$, we can use
the rule $\rulename{abs}$ to form the sequent $\Gamma \sentail \lambda \overline{x}^{\overline{B}} . N : A$.
\end{proof}

\begin{proposition}
\label{prop:safe_iff_elnfsafe}
A term is safe if and only if its $\eta$-long normal form is safe.
\end{proposition}
\begin{proof}
\begin{align*}
  \mbox{(If):}\qquad  \Gamma \sentail \elnf{M}:T &\implies   \Gamma \lsentail \elnf{M}:T &  \mbox{By Lemma \ref{lem:safe_iff_etalong_lsafe} (only if),} \\
  &\implies   \Gamma \sentail M:T &  \mbox{By Lemma \ref{lem:safe_iff_etalong_lsafe} (if).}
\end{align*}
\begin{align*}
  \mbox{(Only if):}\qquad \Gamma \sentail M:T &\implies   \Gamma \lsentail \elnf{M}:T &  \mbox{By Lemma \ref{lem:safe_iff_etalong_lsafe} (only if),} \\
  &\implies \Gamma \sentail \elnf{M}:T &  \mbox{By Lemma \ref{lem:longsafe_imp_safe}. }
\end{align*}
\qedhere
\end{proof}


\subsection*{The type inhabitation problem}

It is well known that the simply-typed lambda calculus corresponds
to intuitionistic implicative logic via the Curry-Howard
isomorphism. The theorems of the logic correspond to inhabited
types, and every inhabitant of a type represents a proof of the
corresponding formula. Similarly, we can consider the fragment of
intuitionistic implicative logic that corresponds to the safe lambda
calculus under the Curry-Howard isomorphism; we call it the
\emph{safe fragment of intuitionistic implicative logic}.

We would like to compare the reasoning power of these
two logics, in other words, to determine which types are inhabited
in the lambda calculus but not in the safe lambda
calculus.\footnote{This problem was raised to our attention by Ugo
dal Lago.}

If types are generated from a single atom $o$, then there is a
positive answer: Every type generated from one atom that is
inhabited in the lambda calculus is also inhabited in the safe
lambda calculus. Indeed, one can transform any unsafe inhabitant $M$
into a safe one of the same type as follows: Compute the eta-long
beta normal form of $M$. Let $x$ be an occurrence of a ground-type
variable in a subterm of the form $\lambda \overline{x} . C[x]$
where $\lambda \overline{x}$ is the binder of $x$ and for some
context $C[-]$ different from the identity (defined as $C[R]\equiv R$ for all $R$). We replace
the subterm $\lambda \overline{x} . C[x]$ by $\lambda \overline{x}.
x$ in $M$. This transformation is sound because both $C[x]$ and $x$
are of the same ground type. We repeat this procedure until the term
stabilizes. This procedure clearly terminates since the size of the
term decreases strictly after each step. The final term obtained is
safe and of the same type as $M$.

This argument cannot be generalized to types generated from multiple
atoms. In fact there are order-$3$ types with only $2$ atoms that
are inhabited in the simply-typed lambda calculus but not in the
safe lambda calculus. Take for instance the order-$3$ type
 $( ((b, a), b),  ((a, b), a),  a)$ for some distinct atoms $a$ and $b$. It is only inhabited by the following family of terms which are all unsafe:
 \begin{align*}
& \lambda f^{((b, a), b)} g^{((a, b), a)} . g (\lambda x_1^a . f (\lambda y_1^b . x_1)) \\
&\lambda f^{((b, a), b)} g^{((a, b), a)} . g (\lambda x_1^a . f (\lambda y_1^b . g (\lambda x_2^a . y_1))) \\
&\lambda f^{((b, a), b)} g^{((a, b), a)} . g (\lambda x_1^a . f (\lambda y_1^b . g (\lambda x_2^a . f (\lambda y_2^b . x_i))) \qquad\mbox{where $i = 1, 2$} \\
&\lambda f^{((b, a), b)} g^{((a, b), a)} . g (\lambda x_1^a . f (\lambda y_1^b . g (\lambda x_2^a . f (\lambda y_2^b . g (\lambda x_3^a . y_i))) \qquad\mbox{where $i = 1, 2$} \\
&\ldots
\end{align*}

Another example is the type of function composition. For any atom
$a$ and natural number $n\in\nat$, we define the types $n_a$ as
follows: $0_a = a$ and $(n+1)_a = n_a \typear a$. Take three
distinct atoms $a$, $b$ and $c$. For any $i,j,k\in\nat$, we write
$\sigma(i,j,k)$ to denote the type
$$\sigma(i,j,k) \equiv (i_a \typear j_b) \typear (j_b \typear k_c) \typear i_a \typear
k_c \ .$$

For all $i$, $j$, $k$, this type is inhabited in the lambda calculus
by the ``function composition term'':
$$\lambda x y z . y (x\,z) \enspace .$$
This term is safe if and only if $i\geq j$ (for the subterm $x\,z$ is safe iff $i = \ord(i_a) = \ord z \geq \ord(x\,z) = \ord(j_b) = j$). In the case $i<j$, the type
$\sigma(i,j,k)$ may still be safely inhabited. For instance
$\sigma(1,3,4)$ is inhabited by the safe term
$$ \lambda x^{1_a \typear 3_b} y^{3_b \typear 4_c} z^{1_c} . y (x (\lambda u^a . u)) \ .$$
The order-$4$ type $\sigma(0,2,0)$, however, is only inhabited by the unsafe term $\lambda x y z . y (x z) $.

Statman showed \cite{Statman1979} that the problem of deciding
whether a type \emph{defined over an infinite number of ground
atoms} is inhabited (or equivalently of deciding validity of an
intuitionistic implicative formula) is PSPACE-complete. The previous
observations suggest that the validity problem for the safe fragment
of implicative logic may not be PSPACE-hard.

    \allowdisplaybreaks

\section{Expressivity}
\label{sec:expressivity}
\subsection{Numeric functions representable in the safe lambda
calculus}

Natural numbers can be encoded in the simply-typed lambda calculus
using the Church Numerals: each $n\in\nat$ is encoded as the term
$\encode{n} = \lambda s^{(o,o)} z^o. s^n z$ of type $I = ((o,o),o,o)$ where
$o$ is a ground type.
We say that a $p$-ary function $f : \nat^p \rightarrow \nat$, for $p \geq 0$,
is represented by a term $F : (I, \ldots, I, I)$ (with $p+1$ occurrences of $I$)
if for all $m_i \in \nat$, $0\leq i\leq p$ we have:
$$F~\encode{m_1} \ldots \encode{m_p} =_\beta \encode{f(m_1,\ldots, m_p)} \enspace .$$
Schwichtenberg \cite{citeulike:622637} showed the following:

\begin{theorem}[Schwichtenberg, 1976]
The numeric functions representable by simply-typed lambda-terms
of type $I\rightarrow \ldots \rightarrow I$ using the Church Numeral
encoding are exactly the multivariate polynomials \emph{extended
with the conditional function}.
\end{theorem}

If we restrict ourselves to safe terms, the representable functions
are exactly the multivariate polynomials:
\begin{theorem}
\label{thm:polychar} The functions representable by safe
lambda-expressions of type $I\rightarrow \ldots \rightarrow I$
are exactly the multivariate polynomials.
\end{theorem}
\begin{proof}
  Natural numbers are encoded as the Church Numerals: $\encode{n} =
  \lambda s z. s^n z$ for each $n \in\nat$.  Addition: For $n,m \in \nat$, $\encode{n+m} =
  \lambda \alpha^{(o,o)} x^o . (\encode{n} \alpha) (\encode{m} \alpha
  x)$. Multiplication: $\encode{n . m} = \lambda \alpha^{(o,o)}
  . \encode{n} (\encode{m} \alpha)$. These terms are all safe, furthermore
  function composition can be safely encoded: take a function $g:\nat^n \rightarrow \nat$
represented by safe term $G$ of type $I^n \typear
I$ and functions $f_1,\ldots,f_n : \nat^p
\rightarrow \nat$ represented by safe terms $F_1,\ldots F_n$
respectively then the composed function $(x_1,\cdots,x_p) \mapsto
g(f_1(x_1,\ldots,x_p),\ldots,f_n(x_1,\ldots,x_p))$ is represented
by the safe term $\lambda c_1\ldots c_p. G (F_1 c_1 \ldots c_p)\ldots
(F_n c_1 \ldots c_p)$.
Hence any multivariate polynomial $P(n_1, \ldots, n_k)$ can be
computed by composing the addition and multiplication terms as
appropriate.

For the converse, let $U$ be a safe lambda-term of type
$I\rightarrow I\rightarrow I$.  The generalization to terms of type
$I^n \rightarrow I$ for every $n\in\nat$ is immediate (they correspond
to polynomials with $n$ variables). By Lemma
\ref{prop:safe_iff_elnfsafe}, safety is preserved by $\eta$-long
normal expansion therefore we can assume that $U$ is in $\eta$-long
normal form.

Let $\mathcal{N}^\tau_\Sigma$ denote the set of safe
$\eta$-long $\beta$-normal terms of type $\tau$ with free variables
in $\Sigma$, and $\mathcal{A}^\tau_\Sigma$ for the set of
$\beta$-normal terms of type $\tau$ with free variables in $\Sigma$
and of the form $\varphi s_1 \ldots s_m$ for some variable
$\varphi:(A_1,\ldots,A_m,o)$ where $m\geq0$ and for all $1\leq i
\leq m$, $s_i \in \mathcal{N}^{A_i}_\Sigma$. Observe that the set
$\mathcal{A}^o_\Sigma$ contains only safe terms but the sets
$\mathcal{A}^\tau_\Sigma$ in general may contain unsafe terms. Let
$\Sigma$ denote the alphabet $\{ x, y : I, z :o, \alpha :
o\typear o \}$.
By an easy reasoning (See the term grammar construction of Zaionc \cite{DBLP:journals/tcs/Zaionc87}), we can derive the following equations inducing a grammar over the set of terminals $\Sigma \union \{\lambda x y \alpha z. , \lambda z. \}$ that generates precisely the terms of $\mathcal{N}^{(I,I,I)}_\emptyset$:
\begin{eqnarray*}
\mathcal{N}^{(I,I,I)}_\emptyset &\rightarrow& \ \lambda x y \alpha z . \mathcal{A}^o_\Sigma \\
\mathcal{A}^o_\Sigma &\rightarrow&\ z \ | \ \mathcal{A}^{(o,o)}_\Sigma \mathcal{A}^o_\Sigma \\
\mathcal{A}^{(o,o)}_\Sigma &\rightarrow&\ \alpha \ |\ \mathcal{A}^I_\Sigma ~\mathcal{N}^{(o,o)}_\Sigma \\
\mathcal{N}^{(o,o)}_\Sigma &\rightarrow& \ \lambda z . \mathcal{A}^{o}_\Sigma \\
\mathcal{A}^I_\Sigma &\rightarrow&\ x \ |\ y \enspace .
\end{eqnarray*}
The key rule is the fourth one: had we not imposed the safety
constraint the right-hand side would instead be of the form $\lambda
w^o . \mathcal{A}^{(o,o)}_{\Sigma\union\{w:o\}}$. Here the safety
constraint imposes to abstract all the ground type variables
occurring freely, thus only one free variable of ground type can
appear in the term and we can choose it to be named $z$ up to
$\alpha$-conversion.

We extend the notion of representability to terms of type $o$,
$(o,o)$ and $I$ with free variables in $\Sigma$ as follows: A
function $f:\nat^2 \rightarrow \nat$ is represented by
(i) $\Sigma \stentail F:o$ if and only if for all $m,n\in\nat$,
$F[\encode{m}, \encode{n}/x,y] =_\beta \alpha^{\overline{f(m,n)}}
z$; (ii) $\Sigma \stentail G:(o,o)$ iff $G[\encode{m},
\encode{n}/x,y] =_\beta \lambda z . \alpha^{\overline{f(m,n)}} z$;
(iii) $\Sigma \stentail H:I$ iff $H[\encode{m}, \encode{n}/x,y]
=_\beta \lambda \alpha z . \alpha^{\overline{f(m,n)}} z$.

We now show by induction on the grammar rules that any term
generated by the grammar represents some polynomial:
\emph{Base case:} The term $x$
and $y$ represent the projection functions $(m,n)\mapsto m$ and
$(m,n)\mapsto n$ respectively. The term $\alpha$ and $z$ represent the constant functions $(m,n)\mapsto 1$ and $(m,n)\mapsto 0$ respectively.
\emph{Step case:} The first and fourth rule are trivial: for $F\in \mathcal{A}^o_\Sigma$, the terms $\lambda z. F$ and $\lambda x y \alpha z. F$ represent the same function as $F$. We now consider the second and third rule. We observe that for $m, p, p' \geq 0$ we have
\begin{align*}
(i)\enspace & {\overline m} \, (\lambda z.\alpha^p z) =_\beta \lambda z . \alpha^{m \cdot p} z; &
(ii)\enspace &(\lambda z.\alpha^p z) (\alpha^{p'} z) =_\beta \alpha^{p + p'} z \enspace .
\end{align*}

Suppose that $F\in \mathcal{A}^I_\Sigma$ and $G\in
\mathcal{N}^{(o,o)}_\Sigma$ represent the functions $f$ and $g$
respectively then by $(i)$, $F G$ represents the
function $f \times g$. If $F\in \mathcal{A}^{(o,o)}_\Sigma$ and
$G\in \mathcal{N}^o_\Sigma$ represent the functions $f$ and $g$ then by $(ii)$,
$F G$ represents the function $f+g$.

Hence $U$ represents some polynomial: for all $m,n\in
\nat$ we have $U~\encode{m}~\encode{n} =_\beta \lambda \alpha z .
\alpha^{p(m,n)} z$ where $p(m,n) = \sum_{0\leq k \leq d} m^{i_k}
n^{j_k}$ for some $i_k,j_k \geq 0$, $d\geq 0$.
\end{proof}

\begin{corollary}
The conditional operator $C:I\rightarrow I\rightarrow I \rightarrow
I$ satisfying:
$$
C~t~y~z \rightarrow_\beta \left\{
                            \begin{array}{ll}
                              y, & \hbox{if $t
\rightarrow_\beta \encode{0}$\enspace ;} \\
                              z, & \hbox{if
$t \rightarrow_\beta \encode{n+1}$\enspace .}
                            \end{array}
                          \right.
$$
is not definable in the simply-typed safe lambda calculus.
\end{corollary}

\begin{example}
The term $\lambda F G H \alpha x . F ( \underline{\lambda y . G
\alpha x} ) (H \alpha x)$ used by Schwichtenberg
\cite{citeulike:622637} to define the conditional operator is unsafe
since the underlined subterm, which is of order $1$, occurs at an
operand position and contains an occurrence of $x$ of order $0$.
\end{example}

\begin{remark} \hfill
\begin{enumerate}[(i)]
\item This corollary tells us that the conditional function is not
definable when numbers are represented by the Church Numerals.
It may still be possible, however, to represent the conditional
function using a different encoding for natural numbers. One way to compensate for the loss of expressivity caused
by the safety constraint is to introduce countably many
domains of representation for natural numbers. Such
a technique is used to represent the predecessor function in
the simply-typed lambda calculus
\cite{DBLP:journals/jacm/FortuneLO83}.

\item The boolean conditional can be represented in the safe
lambda calculus as follows: We encode booleans by terms of type
$B=(o,o,o)$. The two truth values are then represented by
$\lambda x^o y^o.x$ and $\lambda x^o y^o.y$ and the conditional operator is given by
the term $\lambda F^B G^B H^B x^o y^o. F\,(G\,x\,y) (H\,x\,y)$.

\item It is also possible to define a conditional operator behaving like
the conditional operator $C$ in the second-order lambda calculus
\cite{DBLP:journals/jacm/FortuneLO83}: natural numbers are
represented by terms $\overline{n} \equiv \Lambda t.\lambda
s^{t\typear t} z^t.s^n(z)$ of type $J \equiv \Delta t.(t\typear
t) \typear (t \typear t)$ and the conditional is encoded by the
term $\lambda F^J G^J H^J.F~J~(\lambda u^J . G)~H$. Whether this
term is safe or not cannot be answered just yet as we do not
have a notion of safety for second-order typed terms.
\end{enumerate}
\end{remark}

\newcommand{\zaioncencode}{\underline} 

\newcommand{\zaiwordtyp}{\mathbf{B}} 
\newcommand{\closedof}[1]{{\rm Cl}(#1)} 

\newcommand{\openedof}[2]{{\rm Op}(#1,#2)} 

\newcommand\wordnum[1]{\mathbf{#1}} 
\newcommand\safedefset{$\lambda^{{\rm safe}}${\rm def}}

\newcommand\fatlambda{\lambda\kern-0.7em\lambda}
\newcommand\wordapp{\mathop{app}}
\newcommand\wordsub{\mathop{sub}}

\subsection{Word functions definable in the safe lambda calculus.}
Schwichtenberg's result on numeric functions definable in the lambda
calculus was extended to richer structures: Zaionc studied the
problem for word functions, then functions over trees and
eventually the general case of functions over free algebras
\cite{DBLP:journals/tcs/Leivant93,DBLP:journals/apal/Zaionc91,DBLP:conf/aluacs/Zaionc88,DBLP:journals/tcs/Zaionc87,
zaionc:csl94}. In this section we consider the case of word
functions expressible in the safe lambda calculus.
\smallskip

\subsubsection*{Word functions.}  We consider a binary alphabet $\Sigma = \{a,b\}$. The result of this section naturally extends to all finite alphabets. We consider the set $\Sigma^*$ of all words over
$\Sigma$. The empty words is denoted $\epsilon$. We write $|w|$ to
denote the length of the word $w\in\Sigma^*$. For any $k\in \nat$ we
write $\wordnum{k}$ to denote the word $a \ldots a$ with $k$
occurrences of $a$, so that $|\wordnum{k}| = k$. For any $n\geq 1$
and $k\geq 0$, we write $c(n,k)$ for the $n$-ary function
$(\Sigma^*)^n \rightarrow \Sigma^*$ that maps all inputs to the word
$\wordnum{k}$. We consider various word functions. Let $x,y,z$ be words over $\Sigma$:
\begin{enumerate}[$\bullet$]
\item Concatenation $\wordapp : (\Sigma^*)^2 \rightarrow
\Sigma^*$. The word $\wordapp(x,y)$ is the concatenation of $x$ and $y$.
\item Substitution $\wordsub : (\Sigma^*)^3 \rightarrow \Sigma^*$. The word $\wordsub(x,y,z)$ is obtained from $x$ by substituting the word $y$ for all occurrences of $a$ and $z$ for all
occurrences of $b$. Formally:
\begin{align*}
  \wordsub(\epsilon, y, z) &= \epsilon\enspace, \\
  \wordsub(ax, y, z) &= \wordapp(y,\wordsub(x,y,z))\enspace, \\
  \wordsub(bx, y, z) &= \wordapp(z,\wordsub(x,y,z))\enspace .
\end{align*}

\item Prefix-cut $\mathop{cut_a} : \Sigma^* \rightarrow
\Sigma^*$. The word $\mathop{cut_a} x$ is the maximal prefix of $x$ containing only the letter 'a'. Formally:
\begin{align*}
  \mathop{cut_a}(\epsilon) &= \epsilon\enspace, \\
  \mathop{cut_a}(ax) &= \wordapp(a,\mathop{cut_a}(x))\enspace, \\
  \mathop{cut_a}(bx) &= \epsilon\enspace .
\end{align*}
  \item Projections $\pi_k : (\Sigma^*)^n \rightarrow
\Sigma^*$ for $n\geq 1$, $1\leq k \leq n$ defined as $\pi_k(x_1,\ldots,x_k,\ldots,x_n) = x_k$.
  \item Constant functions $\mathop{cst}_w : \Sigma^* \rightarrow
\Sigma^*$ for $w \in \Sigma^*$, mapping constantly onto the word $w$.
\end{enumerate}
Additional operations can be obtained by combining the above functions \cite{DBLP:journals/apal/Zaionc91}:
\begin{enumerate}[$\bullet$]
    \item Prefix-cut $\mathop{cut_b} : \Sigma^* \rightarrow
\Sigma^*$ is defined by $\mathop{cut_b}(x) = \wordsub(\mathop{cut_a}(\wordsub(x,b,a)),b,a)$.

  \item Non-emptiness check $\mathop{\overline{sq}} : \Sigma^* \rightarrow \Sigma^*$ (returns $\wordnum{0}$ if the word is $\epsilon$ and $\wordnum{1}$ otherwise) is defined by $\mathop{\overline{sq}}(x) = \mathop{cut_a}(\wordapp(\wordsub(x,b,b),a)$.

  \item Emptiness check $\mathop{sq} : \Sigma^* \rightarrow \Sigma^*$ is defined by
    $\mathop{sq}(x) = \mathop{\overline{sq}}(\mathop{\overline{sq}}(x))$.

  \item Occurrence check $\mathop{occ_l} : \Sigma^* \rightarrow \Sigma^*$ of the letter $l\in\Sigma$ (returns $\wordnum{1}$ if the word contains an occurrence of $l$ and $\wordnum{0}$ otherwise) is defined by
      $\mathop{occ_l}(x) = \mathop{sq}(\wordsub(x,l,\epsilon))$.
\end{enumerate}

\subsubsection*{Representability}
We consider equality of terms modulo $\alpha$,
$\beta$ and $\eta$ conversion, and we write $M=_{\beta\eta} N$ to
denote this equality. For every simple type $\tau$, we write
$\closedof{\tau}$ for the set of closed terms of type $\tau$ (modulo
$\alpha$, $\beta$ and $\eta$ conversion).

Take the type $\zaiwordtyp = (o\typear o)\typear (o\typear o)\typear
o\typear o$, called \index{type!binary word}\emph{the binary word type}
\cite{DBLP:journals/tcs/Zaionc87}. There is a 1-1 correspondence
between words over $\Sigma$ and closed terms of type $\zaiwordtyp$.
Think of the first two parameters as concatenators for `$a$' and `$b$' respectively, and the third parameter as the constructor for the empty word.
Thus the empty word $\epsilon$ is represented by $\lambda u^{o\typear o} v^{o\typear o} x^o.x$; if $w\in \Sigma^*$ is represented by a term $W \in
\closedof{\zaiwordtyp}$ then $a \cdot w$ is represented by $\lambda u^{o\typear o} v^{o\typear o} x^o. u(W u v x)$ and $b \cdot w$ is represented by $\lambda u^{o\typear o} v^{o\typear o} x^o. v(W u v x)$. For any word $w \in \Sigma^*$ we write $\zaioncencode{w}$ to denote the term representation obtained that way. We say that the
word function $h:(\Sigma^*)^n \rightarrow \Sigma^*$ is
\defname[represented!word function]{represented} by a closed term $H \in \closedof{\zaiwordtyp^n
\typear \zaiwordtyp}$ just if for all $x_1, \ldots, x_n \in
\zaiwordtyp^*$, $H \zaioncencode{x_1} \ldots \zaioncencode{x_n} =_{\beta\eta}
\zaioncencode{h x_1 \ldots x_n}$.

\begin{example}
\label{ex:wordfunc_repres}
  The word functions $\wordapp, \wordsub, \mathop{cut_a}, \mathop{cut_b},
  \mathop{sq}, \mathop{\overline{sq}},
    \mathop{occ_a}, \mathop{occ_b}$ defined above are respectively represented by the following lambda-terms:
\begin{align*}
  {\rm APP} &\equiv \lambda c d u v x.c u v(d u v x), & {\rm SUB} &\equiv \lambda x d e u v x.c(\lambda y.d u v y)(\lambda y . e u v y) x, \\
  {\rm CUT}_a &\equiv \lambda c u v x . c u (\lambda y.x) x,  & {\rm CUT}_b &\equiv \lambda c u v x . c (\lambda y.x) v x, \\
  {\rm SQ} &\equiv \lambda c u v x . c (\lambda y.u x) (\lambda y.u x) x, & \overline{{\rm SQ}} &\equiv \lambda c u v x . c (\lambda y.x) (\lambda y.x) (u x), \\
  {\rm OCC}_a &\equiv \lambda c u v x . c (\lambda y.u x) (\lambda y.y) x, & {\rm OCC}_b &\equiv \lambda c u v x . c (\lambda y.y) (\lambda y.u x) x.
\end{align*}
\end{example}

Zaionc \cite{DBLP:journals/tcs/Zaionc87} showed that
the $\lambda$-definable word functions are generated by a finite base in the following sense:
\begin{theorem}[Zaionc \cite{DBLP:journals/tcs/Zaionc87}]
The set of $\lambda$-definable word functions is the minimal set containing:
\begin{inparaenum}[(i)]
  \item the constant functions;
  \item the projections;
  \item concatenation $\wordapp$;
  \item substitution $\wordsub$;
  \item prefix-cut $\mathop{cut_a}$;
\end{inparaenum}
and closed by composition.
\end{theorem}

The terms representing these basic operations are given in Example \ref{ex:wordfunc_repres}.
We observe that among them, only {\rm APP} and {\rm SUB} are
safe; the other terms are all unsafe because they contain terms of
the form $ N (\lambda y .x)$ where $x$ and $y$ are of the same
order. It turns out that {\rm APP} and {\rm SUB} constitute a base
of terms generating all the functions definable in the safe lambda
calculus as the following theorem states:
\begin{theorem}
\label{thm:wordfunctions_safely_definable}
Let \safedefset\ denote the minimal set containing the following word functions and closed by composition:
\begin{enumerate}[\em(i)]
  \item the projections;
  \item the constant functions;
  \item concatenation $\wordapp$;
  \item substitution $\wordsub$.
\end{enumerate}
The set of word functions definable in the safe lambda calculus is
precisely \safedefset.
\end{theorem}

The proof follows the same steps as Zaionc's proof. The first
direction is immediate: Projections are
represented by safe terms of the form $\lambda x_1 \ldots x_n . x_i$
for some $i\in\{1..n\}$, and constant functions by $\lambda x_1
\ldots x_n . \zaioncencode{w}$ for some $w\in\Sigma^*$. The terms {\rm APP} and {\rm SUB} are safe and represent concatenation and substitution. For
closure by composition: take a function $g:(\Sigma^*)^n \rightarrow \Sigma^*$
represented by safe term $G\in \closedof{\zaiwordtyp^n \typear
\zaiwordtyp}$ and functions $f_1,\ldots,f_n : (\Sigma^*)^p
\rightarrow \Sigma^*$ represented by safe terms $F_1,\ldots F_n$
respectively then the function $$(x_1,\cdots,x_p) \mapsto
g(f_1(x_1,\ldots,x_p),\ldots,f_n(x_1,\ldots,x_p))$$ is represented
by the term $\lambda c_1\ldots c_p. G (F_1 c_1 \ldots c_p)\ldots
(F_n c_1 \ldots c_p)$ which is also safe.
\bigskip

To show the other direction we need to introduce some more definitions.
We will write $\openedof{n}{k}$ to denote the set of open terms
$M$ typable as follows:
$$c_1:\zaiwordtyp, \ldots c_n :\zaiwordtyp, u:(o,o), v:(o,o), x_{k-1}:o, \ldots, x_0 :o \stentail M : o \enspace .$$
Thus we have the following equality (modulo $\alpha$, $\beta$ and
$\eta$ conversions) for $n,k\geq1$:
$$\closedof{\tau(n,k)} = \{ \lambda c_1^\zaiwordtyp \ldots c_n^\zaiwordtyp u^{(o,o)} v^{(o,o)} x_{k-1}^o \ldots x_0^o . M \ | \ M \in \openedof{n}{k}  \} $$
writing $\tau(n,k)$ as a shorthand for the type
$\zaiwordtyp^n \typear (o,o)^2\typear o^k\typear o$. We generalize the notion of representability to terms
of type $\tau(n,k)$ as follows:
\begin{definition}[Function pair representation]
A closed term $T\in\closedof{\tau(n,k)}$ \defname{represents the pair of functions}
$(f,p)$ where $f:(\Sigma^*)^n \rightarrow \Sigma^*$ and $p:(\Sigma^*)^n \rightarrow \{\wordnum{0}, \ldots, \wordnum{k-1}\}$ if for all $w_1,\ldots,w_n\in\Sigma^*$
and for every $i\in \{0\ldots,k-1\}$ we have:
$$
T \zaioncencode{w_1} \ldots \zaioncencode{w_n} =_{\beta\eta} \lambda u v x_{k-1}\ldots x_0 . \zaioncencode{f(w_1,\ldots,w_n)} u v x_{|p(w_1,\ldots,w_n)|} \enspace .
$$
By extension we will say that an \emph{open} term $M$ from $\openedof{n}{k}$
represents the pair $(f,p)$
 just if $M[\zaioncencode{w_1}\ldots \zaioncencode{w_n}/c_1\ldots c_n] =_{\beta\eta} \zaioncencode{f(w_1,\ldots,w_n)} u v x_{|p(w_1,\ldots,w_n)|}$.
\end{definition}

We will call \defname[safe!pair]{safe pair} any pair of functions of the form
$(w,c(n,i))$ where $0\leq i\leq k-1$ and $w$ is an $n$-ary function
from \safedefset.

\begin{theorem}[Characterization of the representable pairs]
\label{thm:zaionc_pair_characterization_safe} The function pairs
representable in the safe lambda calculus are precisely the safe
pairs.
\end{theorem}

\begin{proof}
  (Soundness). Take a pair $(w,c(n,i))$ where
  $0\leq i\leq k-1$ and $w$ is an $n$-ary function from \safedefset.
  As observed earlier, all the functions from \safedefset\ are representable
  in the safe lambda calculus: Let $\zaioncencode{w}$ be the representative of $w$.
  The pair $(w,c(n,i))$ is then represented by the term
  $ \lambda c_1 \ldots c_n u v x_{k-1} \ldots x_0 . \zaioncencode{w} c_1\ldots c_n u v x_i$.
\smallskip

(Completeness) It suffices to consider safe $\beta$-$\eta$-long
normal terms from $\openedof{n}{k}$ only. The result then
follows immediately for every safe term in $\closedof{\tau(n,k)}$. The
subset of $\openedof{n}{k}$ consisting of $\beta$-$\eta$-long
normal terms is generated by the following grammar \cite{DBLP:journals/tcs/Zaionc87}:
\begin{eqnarray*}
  (\alpha_i^k) &R^k &\rightarrow\ x_i \\
  (\beta^k) && \quad|\  u R^k \\
  (\gamma^k) && \quad|\  v R^k \\
  (\delta^k_j) && \quad|\  c_j\ (\overbrace{\lambda z^k. R^{k+1}[z^k,x_0,\ldots, x_{k-1}/x_0,x_1, \ldots, x_k]}^{Q^k(R^{k+1})}) \\
  && \quad\  \quad \ (\lambda z^k. R^{k+1}[z^k,x_0,\ldots, x_{k-1}/x_0,x_1, \ldots, x_k]) \\
  && \quad\  \quad \ R^k
\end{eqnarray*}
for $k\geq 1$, $0\leq i< k$, $0\leq j\leq n$. The notation $M[\ldots/\ldots]$ denotes the usual simultaneous substitution. The non-terminals are
$R^k$ for $k\geq1$ and the set of terminals is $\{ z^k, \lambda z^k
\ |\ k\geq 1\} \union \{ x_i ~| i\geq 0 \} \union \{ c_1, \ldots,
c_n, u, v \}$.

The name of each rule is indicated in parenthesis. We identify a rule name with the right-hand side of the
rule, thus $\alpha_i^k$ belongs to $\openedof{n}{k}$,
$\beta^k$ and $\gamma^k$ are functions from $\openedof{n}{k}$ to
$\openedof{n}{k}$, and $\delta^k_j$ is a function from
$\openedof{n}{k+1} \times \openedof{n}{k+1} \times \openedof{n}{k}$
to $\openedof{n}{k}$.

We now want to characterize the subset consisting of all \emph{safe}
terms generated by this grammar. The term $\alpha_i^k$ is always
safe; $\beta^k(M)$ and $\gamma^k(M)$ are safe if and only if $M$ is;
and  $\delta^k_j(F,G,H)$ is safe if and only if $Q^k(F)$, $Q^k(G)$
and $H$ are safe. The free variables of $Q^k(F)$
belong to $\{ c_1, \ldots c_n, u, v, x_0,\ldots x_{k}\}$ thus they have order greater than $\ord{z}$ except the $x_i$s which have the same order as $z$. Hence since the $x_i$s are not abstracted together with $z$ we have that $Q^k(F)$ is safe if and only if $F$
is safe and the variables $x_0\ldots x_k$ do not appear free in
$F[z^k,x_0,\ldots, x_{k-1}/x_0,x_1, \ldots, x_k]$, or equivalently if the variables $x_1\ldots x_k$ do not appear free in $F$. Similarly, $Q^k(G)$ is safe if and only if $G$ is safe and
the variables $x_1\ldots x_k$ do not appear free in $G$.

We therefore need to identify the subclass of terms generated by the non-terminal $R^k$ which are safe and which do not have any free occurrence of variables in $\{x_1 \ldots x_{k-1}\}$. By imposing this requirement to the rules of the previous grammar we obtain the following specialized grammar characterizing the desired subclass:
\begin{eqnarray*}
  (\overline\alpha_0^k) &\overline R^k &\rightarrow\ x_0 \\
  (\overline\beta^k) && \quad|\  u \overline R^k \\
  (\overline\gamma^k) && \quad|\  v \overline R^k  \\
  (\overline\delta^k_j) && \quad|\  c_j\ (\lambda z^k. \overline R^{k+1}[z^k/x_0]) \ (\lambda z^k. \overline R^{k+1} [z^k/x_0]) \ \overline R^k \enspace .
\end{eqnarray*}
For every term $M$, $Q^k(M)$ is safe if and only if $M$ can be
generated from the non-terminal $\overline R^k$. Thus the subset of
$\closedof{\tau(n,k)}$ consisting of safe beta-normal terms is given
by the grammar:
\begin{eqnarray*}
  (\widetilde\pi^k) &\widetilde S &\rightarrow \lambda c_1 \ldots c_n u v x_{k-1} \ldots x_0 . \widetilde R^k \\
  (\widetilde\alpha_i^k) &\widetilde R^k &\rightarrow\ x_i \\
  (\widetilde\beta^k) && \quad|\  u \widetilde R^k \\
  (\widetilde\gamma^k) && \quad|\  v \widetilde R^k \\
  (\widetilde\delta^k_j) && \quad|\  c_j\ (\lambda z^k. \overline{R^{k+1}}[z^k/x_0]) \ (\lambda z^k. \overline{R^{k+1}}[z^k/x_0]) \ \widetilde R^k \enspace .
\end{eqnarray*}

To conclude the proof it thus suffices to show that every term generated by this grammar (starting with the non-terminal $\widetilde S$) represents a safe pair.

We proceed by induction and show that the non-terminal $\overline
R^k$ generates terms representing pairs of the form $(w,c(n,0))$
while non-terminals $\widetilde S$ and $\widetilde R^k$ generate
terms representing pairs of the form $(w,c(n,i))$ for $0 \leq i<k$
and $w \in$\safedefset.

\emph{Base case:} The term $\overline\alpha_0^k$ represents the safe pair $(c(n,0),c(n,0))$ while
$\widetilde\alpha_i^k$ represents the safe pair
$(c(n,0),c(n,i))$. \emph{Step case:} Suppose $T\in
\openedof{n}{k}$ represents
 a pair $(w,p)$.  Then $\overline\beta^k(T)$ and
 $\widetilde\beta^k(T)$ represent the pair
 $(\wordapp(a,w),p)$; $\overline\gamma^k(T)$ and
 $\widetilde\gamma^k(T)$ represent the pair
 $(\wordapp(b,w),p)$; and $\overline\pi^k(T) \in \closedof{\tau(n,k)}$ represents the pair $(w,p)$. Now suppose that $E$, $F$ and $G$ represent the pairs
 $(w_e,c(n,0))$, $(w_f,c(n,0))$ and $(w_g,c(n,i))$ respectively.
 Then we have:
 \begin{alignat*}{2}
   \widetilde \delta^k_j (E,F,G) &[\zaioncencode{w_1}\ldots \zaioncencode{w_n}/c_1\ldots c_n] \\
   &= \zaioncencode{w_j}\  (\lambda z^k. E[z^k/x_0])[\zaioncencode{w_1}\ldots \zaioncencode{w_n}/c_1\ldots c_n] \\
       & \qquad\quad (\lambda z^k. F[z^k/x_0])[\zaioncencode{w_1}\ldots \zaioncencode{w_n}/c_1\ldots c_n] \\
       & \qquad\quad  G[\zaioncencode{w_1}\ldots \zaioncencode{w_n}/c_1\ldots c_n] \\
   &=_{\beta\eta} \zaioncencode{w_j}\  (\lambda z^k. E[\zaioncencode{w_1}\ldots \zaioncencode{w_n}/c_1\ldots c_n][z^k/x_0]) \\
       & \qquad\qquad (\lambda z^k. F[\zaioncencode{w_1}\ldots \zaioncencode{w_n}/c_1\ldots c_n][z^k/x_0]) \\
       & \qquad\qquad  (\zaioncencode{w_g(w_1\ldots w_n)}~u~v~x_i) \hspace{4cm}\mbox{$G$ represents $(h,c(n,i))$}\\
   &=_{\beta\eta} \zaioncencode{w_j}\  (\lambda z^k. (\zaioncencode{w_e(w_1\ldots w_n)}~u~v~x_0)[z^k/x_0]) \hspace{2cm}\mbox{$E$ represents $(f,c(n,0))$} \\
       & \qquad\qquad (\lambda z^k. (\zaioncencode{w_f(w_1\ldots w_n)}~u~v~x_0)[z^k/x_0]) \hspace{1.8cm}\mbox{$F$ represents $(g,c(n,0))$} \\
       & \qquad\qquad  (\zaioncencode{w_g(w_1\ldots w_n)}~u~v~x_i)\\
   &=_{\beta\eta} \zaioncencode{w_j}\  (\lambda z^k. \zaioncencode{w_e(w_1\ldots w_n)}~u~v~z^k) \\
       & \qquad\qquad (\lambda z^k. \zaioncencode{w_f(w_1\ldots w_n)}~u~v~z^k) \\
       & \qquad\qquad (\zaioncencode{w_g(w_1\ldots w_n)}~u~v~x_i)\\
   &=_\eta \zaioncencode{w_j}\  (\zaioncencode{w_e(w_1\ldots w_n)}~u~v)  \ (\zaioncencode{w_f(w_1\ldots w_n)}~u~v) \  (\zaioncencode{w_g(w_1\ldots w_n)}~u~v~x_i)\\
   &=_{\beta\eta}  \zaioncencode{w}~u~v~ x_i
 \end{alignat*}
where the word function $w$ is defined as
$$w: w_1,\ldots,w_n \mapsto \wordapp(\wordsub(w_j,w_e(w_1,\ldots,w_n),w_f(w_1,\ldots,w_n)),w_g(x_1,\ldots,w_n)) \enspace .$$
  Hence $\widetilde \delta^k_j (E,F,G)$ represents the pair $(w,c(n,i))$.

  The same argument shows that if $E$, $F$ and $G$ all represent safe pairs
then so does $\overline \delta^k_j (E,F,G)$.
\end{proof}

Theorem \ref{thm:wordfunctions_safely_definable} is obtained by
instantiating Theorem \ref{thm:zaionc_pair_characterization_safe}
with terms of types $\tau(n,1) = I^n\typear I$: every
closed safe term of this type represents some $n$-ary function from
\safedefset.

\subsection{Representability of functions over other structures}\hfill

There is an isomorphism between binary trees and closed terms of
type $\tau =(o\typear o\typear o) \typear o \typear o$. Thus a
closed term of type $\tau\typear\tau \typear \ldots \typear \tau $
represents an $n$-ary function over trees. Zaionc gave a
characterization of the set of tree functions representable in the
simply-typed lambda calculus \cite{DBLP:conf/aluacs/Zaionc88}: It is
precisely the minimal set containing constant functions, projections
and closed under composition and limited primitive recursion. Zaionc
showed that the same characterization holds for the general case of
functions expressed over (different) free algebras
\cite{DBLP:journals/apal/Zaionc91,zaionc:csl94} (they are again given by the
minimal set containing constant functions, projections and closed
under composition and limited primitive recursion). This result subsumes
Schwichtenberg's result on definable numeric functions as well as
Zaionc's own results on definable word and tree functions.

We have seen that constant functions, projections and composition can be encoded by safe terms. Limited primitive recursion, however, cannot be encoded in the safe lambda calculus
(It can be used to define the conditional operator and the $\mathop{cut_a}$ word function).
We expect an appropriate restriction to limited recursion to characterize the functions
over free algebras representable in the safe lambda calculus.

    \newcommand\bigo{\mathcal{O}} 
\newcommand\booltype{\mathsf{B}}
\newcommand\towerexp[2]{\exp_{#1}(#2)}

\section{Complexity of the safe lambda calculus}
This section is concerned with the complexity of the
beta-eta equivalence problem for the safe lambda calculus:
Given two safe lambda-terms, are they equivalent up to $\beta\eta$-conversion?

\subsection{Statman's result}

Let $\towerexp{h}{m}$ denote the tower-of-exponential function
defined by induction as $\towerexp{0}{m} = m$ and $\towerexp{h+1}{m} =
2^{\towerexp{h}{m}}$.
A program is \defname{elementary recursive} if its run-time can be bounded by
$\towerexp{K}{n}$ for some constant $K$ where $n$ is the length of
the input.

We recall the definition of finite type
theory. We define $\mathcal{D}_0 =
\{\mathbf{true},\mathbf{false}\}$ and $\mathcal{D}_{k+1}
=\mathscr{P}(\mathcal{D}_k)$ (\ie, the powerset of $\mathcal{D}_k$).
For $k\geq0$, we write $x^k$, $y^k$ and
$z^k$ to denote variables ranging over $\mathcal{D}_k$. Prime
formulae are $x^0$, $\mathbf{true}\in y^1$, $\mathbf{false}\in y^1$,
and  $x^k \in y^{k+1}$. Formulae are built up from prime formulae
using the logical connectives $\zand$,$\zor$,$\rightarrow$,$\neg$
and the quantifiers $\forall$ and $\exists$. Meyer showed that
deciding the validity of such formulae requires nonelementary time
\cite{Meyer1974}.

A famous result by Statman  states that deciding the
$\beta\eta$-equality of two first-order typable lambda-terms is not
elementary recursive \cite{Statman:1979:TLE}. The proof proceeds by
encoding the Henkin quantifier elimination of type theory in the
simply-typed lambda calculus and by appealing to Meyer's result \cite{Meyer1974}. Simpler proofs have subsequently been given: one by Mairson \cite{mairson1992spt} and another by Loader
\cite{Loader1998}. Both proceed by encoding the Henkin quantifier
elimination procedure in the lambda calculus, as in the original
proof, but their use of list iteration to implement quantifier
elimination makes them much easier to understand.

It turns out that all these encodings rely on unsafe terms:
Statman's encoding uses the conditional function $\sf sg$ which is
not definable in the safe lambda calculus
\cite{blumong:safelambdacalculus}; Mairson's encoding uses unsafe
terms to encode both quantifier elimination and set membership, and
Loader's encoding uses unsafe terms to build list iterators. We are
thus led to conjecture that finite type theory (see definition in
Sec.~\ref{sec:mairsonenc}) is intrinsically unsafe in the sense
that every encoding of it in the lambda calculus is necessarily
unsafe. Of course this conjecture does not rule out the possibility
that another non-elementary problem is encodable in the safe lambda
calculus.

\subsection{Mairson's encoding}
\label{sec:mairsonenc}
We refer the reader to Mairson's original paper \cite{mairson1992spt} for a detailed account of his encoding. We show here why Mairson's encoding does not work in the safe lambda calculus. We then introduce a variation that eliminates some of the unsafety. Although the resulting encoding does not suffice to interpret type theory in the safe lambda calculus, it enables another interesting encoding: that of the True Quantifier Boolean Formula (TQBF) problem. This implies that deciding beta-eta equality of safe terms is PSPACE-hard.

\subsubsection{Sources of unsafety}
In Mairson's encoding, boolean values are encoded by terms of type
$\booltype = \sigma \typear \sigma\typear \sigma$ for some type
$\sigma$, and variables of order $k \geq 0$ are encoded by terms of
type $\Delta_k$ defined as $\Delta_0 \equiv \booltype$ and
$\Delta_{k+1} \equiv (\Delta_k \typear \tau \typear \tau)\typear \tau \typear
\tau$ for any type $\tau$. Using this encoding, unsafety manifests
itself in three different places:
\begin{enumerate}[(i)]
  \item \emph{Set membership:} The prime formula ``$x^k \in y^{k+1}$'' is encoded by a term-in-context of the form
      \begin{equation} x: \Delta_k, y:\Delta_{k+1} \stentail y (\lambda z^{\Delta_k} . M(x,z))\,F : \Delta_k \typear \Delta_{k+1} \typear \Delta_0 \label{eqn:setmembership}\end{equation}
for some term $F$ and term $M(x,z)$ containing free occurrences of $x$ and $z$.
This is unsafe because the free occurrence of $x$ in $M(x,z)$ is not abstracted together with $z$.

\item \emph{Quantifier elimination} is implemented using a list iterator $\mathbf{D}_{k+1}$ of type $\Delta_{k+2}$ which acts like the {\tt foldr} function (from functional programming) over the list of all elements of $\mathcal{D}_k$.
Thus nested quantifiers in the formula are encoded by nested list iterations. This can be source of unsafety, for instance the formula ``$\forall x^0 . \exists y^0 . x^0 \zor y^0$'' is encoded as $$\stentail \mathbf{D}_0 (\lambda
x^{\Delta_0}. AND (\mathbf{D}_0 (\lambda y^{\Delta_0}. OR
(\underline{x} \zor y)) F))\ T : \booltype$$
for some terms $AND$, $OR$, $F$ and $T$ and where the type
$\tau$ is instantiated as $\booltype$. This term is unsafe due to
the underlined occurrence which is unsafely bound.

More generally, nested binding will be encoded safely if and only if every variable $x$ in the formula is bound by the first quantifier $\exists z$ or
$\forall z$ satisfying $\ord{z} \geq \ord{x}$ in the path to the root of the formula AST. So for example if set-membership
were safely encodable then the interpretation of ``$\forall x^k
. \exists y^{k+1} . x^k \in y^{k+1}$'' would be unsafe whereas that of ``$\forall y^{k+1} . \exists x^k . x^k \in y^{k+1}$''
would be safe.

\item \emph{Elements of the type hierarchy.} The base set $\mathcal{D}_0$ of booleans is represented by a safe term $\mathbf{D}_0$ of type $\Delta_0$. Higher-order sets $\mathcal{D}_k$ for $k\geq 1$ are represented by unsafe terms $\mathbf{D}_k$: they are constructed from $\mathbf{D}_0$ using a powerset construction that is unsafe.
\end{enumerate}
\bigskip

The second source of unsafety can be easily overcome, the idea is as follows. We introduce multiple domains of representation for a given formula. An element of $\mathcal{D}_k$ is thereby represented by countably many terms of type $\Delta_k^n$ where $n\in\nat$ indicates the level of the domain of representation. The type $\Delta_k^n$ is defined in such a way that its order strictly increases as $n$ grows. Furthermore, there exists a term that can lower the domain of representation of a given term. Thus each formula variable can have a different domain of representation, and since there are infinitely many such domains, it is always possible to find an assignment of representation domains to variables such that the resulting encoding term is safe.

There is no obvious way to eliminate unsafety in the two other cases however.
For instance in the case of set-membership, Mairson's encoding (\ref{eqn:setmembership}) could be made safe by appealing to a term that changes the domain of representation of an encoded higher-order value of the type-hierarchy. Unfortunately, such transformation is intrinsically unsafe!
\smallskip

In the following paragraphs we present in detail a variation over Mairson's encoding in which quantifier elimination is safely encoded.

\subsubsection{Encoding basic boolean operations}

Let $o$ be a base type and define the family of types $\sigma_0
\equiv o$, $\sigma_{n+1} \equiv \sigma_n\typear\sigma_n$ satisfying
$\ord{\sigma_n} = n$. Booleans are encoded over domains $\booltype_n
\equiv \sigma_n\typear o\typear o\typear o$ for $n\geq0$, each type
$\booltype_n$ being of order $n+1$. We write $\underline{i}_{n+1}$
to denote the term $\lambda x^{\sigma_n}.x$ of type $\sigma_{n+1}$ for
$n\geq0$. The truth values $\mathbf{true}$ and $\mathbf{false}$ are
represented by the following terms parameterized by $n \in \nat$:
\begin{align*}
  T^n &\equiv \lambda u^{\sigma_n} x^o y^o .x : \booltype_{n}\\
  F^n &\equiv \lambda u^{\sigma_n} x^o y^o .y : \booltype_{n} \enspace .
\end{align*}
Clearly these terms are safe. Moreover the following relations hold
for all $n,n'\geq 0$:
\begin{align*}
  \lambda u^{\sigma_{n'}} . T^{n+1}~\underline{i}_{n+1}  &\betared  T^{n'} \\
  \lambda u^{\sigma_{n'}} . F^{n+1}~\underline{i}_{n+1}  &\betared  F^{n'} \enspace .
\end{align*}
It is then possible to change the domain of representation of a Boolean value from a higher-level to another arbitrary level using the conversion term:
$$ \mathbf{C}^{n+1\mapsto n'}_0 \equiv \lambda m^{\booltype_{n+1}} u^{\sigma_{n'}}. m~\underline{i}_{n+1} : \booltype_{n+1} \typear \booltype_{n'}$$
so that if a term $M$ of type $\booltype_n$, for $n\geq1$, is beta-eta convertible to $T^n$ (resp.\ $F^n$) then $\mathbf{C}^{n\mapsto n'}_0~M$ of type $\booltype_{n'}$ is beta-eta convertible to $T^{n'}$ (resp.\ $F^{n'}$).

Observe that although $\mathbf{C}^{n+1\mapsto n'}_0$ is safe for all $n,n'\geq 0$, if we apply a variable to it then the resulting term-in-context
$$ x:B_{n+1} \stentail \mathbf{C}_0^{n+1\mapsto n'}~x : B_{n}$$
is safe if and only if $\ord{B_{n+1}}\geq\ord{B_{n'}}$, that is to say if and only if the transformation decreases the domain of representation of $x$.

Boolean functions are encoded by the following closed safe terms parameterized by $n$:
\begin{align*}
AND^n &\equiv \lambda p^{\booltype_n} q^{\booltype_n} u^{\sigma_n} x^o y^o . p~u~(q~u~x~y)~y : \booltype_n\typear\booltype_n\typear\booltype_n \\
OR^n &\equiv \lambda p^{\booltype_n} q^{\booltype_n} u^{\sigma_n} x^o y^o . p~u~x~(q~u~x~y) : \booltype_n\typear\booltype_{n}\typear\booltype_n \\
NOT^n &\equiv \lambda p^{\booltype_n} u^{\sigma_n} x^o \lambda y^o . p~u~y~x : \booltype_n\typear\booltype_n\typear\booltype_n
\enspace .
\end{align*}

\subsubsection{Coding elements of the type hierarchy}
For every $n\in\nat$ we define the hierarchy of type $\Delta_k^n$ as
follows: $\Delta_0^n \equiv \booltype_n$ and $\Delta_{k+1}^n \equiv
{\Delta_k^n}^*$ where for a given type $\alpha$, $\alpha^* = (\alpha
\typear \tau \typear \tau)\typear \tau \typear \tau$ for \emph{any} type $\tau$. We encode an occurrence $x^k$ of a \emph{formula variable} by a \emph{term variable} $x^k$ of type $\Delta_{k}^n$ for some level of domain representation
$n\in\nat$. Following Mairson's  encoding, each set $\mathcal{D}_k$
is represented by a list $\mathbf{D}_k^n$ consisting of all its
elements:
\begin{align*}
\mathbf{D}_0^n &\equiv \lambda c^{\booltype_n \typear \tau \typear \tau} e^\tau . c~T^n~(c~F^n~e) : \Delta_1^n \\
\mathbf{D}_{k+1}^n &\equiv powerset_{\Delta_k^n}\,\mathbf{D}_k^n : \Delta_{k+2}^n
\end{align*}
where
\begin{align*}
  powerset_\alpha &\equiv \lambda {A^*}^{(\alpha \typear \alpha^{**} \typear \alpha^{**}) \typear \alpha^{**} \typear \alpha^{**}}. \\
&\qquad  A^*~double_\alpha~(\lambda c^{\alpha^* \typear \tau\typear \tau} b^\tau . c~(\lambda c'^{\alpha\typear \tau\typear \tau} b'^\tau.b')~b)\\
 &: ((\alpha \typear \alpha^{**} \typear \alpha^{**}) \typear \alpha^{**} \typear \alpha^{**})\typear \alpha^{**} \\
  double_\alpha &\equiv \lambda x^\alpha~l^{(\alpha^* \typear \tau\typear \tau)\typear \tau\typear \tau}~ c^{\alpha^*\typear \tau\typear \tau}~b^\tau. \\
  & \qquad \qquad l(\lambda e^{\alpha^*}.c~(\lambda c'^{\alpha\typear \tau\typear \tau}~ b'^\tau.c'~\underline{x}~(e~c'~b')))(l~c~b)\\
 &: \alpha \typear \alpha^{**} \typear \alpha^{**} \enspace .
\end{align*}
(In the definition of $\mathbf{D}_{k+1}^n$, to see why it is possible
to apply $powerset_{\Delta_k^n}$ and $\mathbf{D}_k^n$ one needs to understand that the term $\mathbf{D}_k^n$ is of type $\Delta^n_{k+1}$ \emph{polymorphic in $\tau$}. The application can thus be typed by taking $\tau \equiv \Delta^n_{k+2}$ in the term
$\mathbf{D}_k^n$.)

Observe that the term $double$ is unsafe because the underlined variable occurrence $x$ is not bound together with $c'$.
Consequently for all $n\geq0$, $\mathbf{D}_0^n$ is safe and $\mathbf{D}_k^n$ is unsafe for all $k>0$.

\subsubsection{Quantifier elimination}
Terms of type $\Delta_{k+1}^n$ are now used as iterators over lists of elements of type $\Delta_k^n$ and we set $\tau \equiv \booltype_n$ in the type $\Delta_{k+1}^n$ in order to iterate a level-$n$ Boolean function. Since $\ord{\Delta_k^n} \geq \ord{\booltype_n}$ for all $n$, all the instantiations of the terms $\mathbf{D}_k^n$ will be safe
 (although the terms $\mathbf{D}_k^n$ themselves are not safe for $k>1$).
 Following \cite{mairson1992spt}, quantifier elimination interprets the formula $\forall x^k.\Phi(x^k)$ as the iterated conjunction
$\mathbf{C}_0^{n\mapsto 0} \left( \mathbf{D}_k^n(\lambda x^{\Delta_k^n}.AND^n(\hat\Phi\,x))\,T^n \right)$ where $\hat\Phi$ is the interpretation of $\Phi$
and $n$ is the representation level chosen for the variable $x^k$. Similarly we interpret $\exists x^k.\Phi(x^k)$  by the iterated disjunction $\mathbf{C}_0^{n\mapsto 0} \left( \mathbf{D}_k^n(\lambda x^{\Delta_k^n}.AND^n(\hat\Phi\,x))\,T^n\right)$.

\subsubsection{Encoding the formula}

Given a formula of type theory, it is possible to encode it in the lambda calculus by inductively applying the above encodings of boolean operations and quantifiers on the formula; each variable occurrence in the formula being assigned some domain of representation.

We now show that there exists an assignment of representation domains for each variable occurrence such that the resulting term is safe. Let $x^{k_p}_p \ldots x^{k_1}_1$ for $p\geq1$ be the list of variables appearing in the formula, given in order of appearance of their binder in the formula (\ie,~$x^{k_p}_p$ is bound by the leftmost binder). We fix the domain of representation of each variable as follows. The right-most variable $x^{k_1}_1$ is encoded in the domain $\Delta^0_{k_1}$; and if for $1\leq i< p$ the domain of representation of $x^{k_i}_i$ is $\Delta^l_{k_l}$ then the domain of representation of $x^{k_{i+1}}_{i+1}$ is defined as
$\Delta^{l'}_{k_{i+1}}$ where $l'$ is the smallest natural number such that $\ord{\Delta^{l'}_{k_{i+1}}}$ is strictly greater than $\ord{\Delta^{l}_{k_i}}$.

This way, since variables that are bound first have higher order, variables that are bound in  nested list-iterations---corresponding to nested quantifiers in the formula---are guaranteed to be safely bound.

\begin{example}
The formula  $\forall x^0 . \exists y^0 . x^0 \zor y^0$, which is
encoded by an unsafe term in Mairson's encoding, is represented in
our encoding by the safe term
 $$\sentail \mathbf{C}_0^{1\mapsto 0} \left( \mathbf{D}_0^1~(\lambda x^{\Delta_0^1}. AND^0 ( \mathbf{D}_0^0 ~(\lambda y^{\Delta_0^0}. OR^0 (OR^0~(\mathbf{C}_0^{1\mapsto 0}~x)~y))~F^0))~T^1 \right) : \booltype_0 \enspace .$$
\end{example}

\subsubsection{Set-membership}
To complete the interpretation of prime formulae, we need to show how to encode set membership. Unfortunately, the introduction of multiple domains of representation does not permit us to completely eliminate the unsafety of Mairson's encoding of set membership.

Indeed, adapting Mairson's encoding of set membership requires the ability to perform conversion of domains of representation for higher-order sets (not only for Boolean values). The conversion term $\mathbf{C}^{n+1\mapsto n'}_0$ can be generalized to higher-order sets as follows:
$$ \mathbf{C}^{n\mapsto n'}_{k+1} \equiv \lambda m^{\Delta_{k+1}^n} u^{\Delta_k^n\typear \tau\typear \tau} v^\tau. m (\lambda z^{\Delta_k^n} w^\tau . \underline{u (\underline{\mathbf{C}^{n\mapsto n'}_k z}) w}) v : \Delta_{k+1}^n \typear \Delta_{k+1}^{n'}$$
where $k\geq 0$. Unfortunately this term is safe if and only if $n=n'$ (The largest underlined subterm is safe just when $n\geq n'$ and the other
underline subterm is safe just when $n'\geq n$). Hence at higher-orders, all the non-trivial conversion terms are unsafe.
\smallskip

If the terms $\mathbf{C}^{n\mapsto n'}_{k+1}$, $k\geq 0$, $n\neq n'$ were safely representable then the encoding would go as follows: We set $\tau \equiv \booltype_0$ in the types  $\Delta_{k+1}^n$ for all $n,k\geq 0$ in order to iterate a level-$0$ Boolean function.
Firstly, the formulae ``$\mathbf{true} \in y^1$'' and ``$\mathbf{false} \in y^1$'' can be encoded by the safe terms $y^1 (\lambda x^0 . OR^0~x^0) F^0$ and $y^1 (\lambda x^0. OR^0(NOT^0~x^0)) F^0$ respectively.
For the general case ``$x^k\in y^{k+1}$''
 we proceed as in Mairson's proof \cite{mairson1992spt}: we introduce lambda-terms encoding set equality, set membership and subset tests, and we further parameterize these encodings by a natural number $n$.
\begin{align*}
member_{k+1}^{n+1} &\equiv \lambda x^{\Delta_k^{n+1}} y^{\Delta_{k+1}^{n+1}}.
(\mathbf{C}_{k+1}^{n+1\mapsto n}~y)~(\lambda z^{\Delta_k^n} . OR^0 (eq_k^n~(\mathbf{C}^{n+1\mapsto n}_k~x)~z))~F^0\\
  & : \Delta_k^{n+1} \typear \Delta_{k+1}^{n+1} \typear \booltype_0
\\
subset_{k+1}^n &\equiv \lambda x^{\Delta_{k+1}^n} y^{\Delta_{k+1}^n}.
x~(\lambda x^{\Delta_k^n} . AND^0 (member_{k+1}^n~x~y))~T^0 \\
  & : \Delta_{k+1}^n \typear \Delta_{k+1}^n \typear\booltype_0
\\
eq_0^n &\equiv \lambda x^{\booltype_n} .\lambda y^{\booltype_n}. \mathbf{C}_0^{n\mapsto 0}~ \left(OR^n (AND^n~x~y) (AND^n (NOT^n~x)(NOT^n~y))\right) \\
 &: \booltype_n \typear \booltype_n \typear\booltype_0
\\
eq_{k+1}^n &\equiv \lambda x^{\Delta_{k+1}^n}~ y^{\Delta_{k+1}^n}.
   (\lambda op^{\Delta_{k+1}^n\typear\Delta_{k+1}^n\typear\booltype_0}. AND^0 (op~x~y)(op~y~x))~subset_{k+1}^n \\
  & : \Delta_{k+1}^n \typear \Delta_{k+1}^n \typear \booltype_0 \enspace .
\end{align*}
The variables in the definition of $eq_{k+1}^n$ and $subset_{k+1}^n$
are safely bounds. Moreover, the occurrence of $x$ in
$member_{k+1}^{n+1}$ is now safely bound---which was not the case in
Mairson's original encoding---thanks to the fact that the
representation domain of $z$ is lower than that of $x$. The formula
$x^k\in y^{k+1}$ can then be encoded as
$$x:\Delta_k^n, y:\Delta_{k+1}^{n'}\stentail member_{k+1}^{u}~ (\mathbf{C}_k^{n\mapsto u}~x)~(\mathbf{C}^{n'\mapsto u}_{k+1} ~y) : \booltype_0$$
for some $n,n'\geq 2$ and $u = \min(n,n')+1$.

Unfortunately this encoding is not completely safe because, as mentioned before, the conversion term $\mathbf{C}_k^{n\mapsto u}$ is unsafe for
$k\geq1$, $n\neq u$. We conjecture that the set-membership function is intrinsically unsafe.

\subsection{PSPACE-hardness}
We observe that instances of the True Quantified Boolean Formulae
satisfaction problem (TQBF) are special instances of the decision
problem for finite type theory. These instances correspond to
formulae in which set membership is not allowed and variables are
all taken from the base domain $\mathcal{D}_0$. As we have shown in
the previous section, such restricted formulae can be safely encoded
in the safe lambda calculus. Therefore since TQBF is
PSPACE-complete we have:
\begin{theorem}
\label{thm:pspacehardness}
 Deciding $\beta\eta$-equality of two safe lambda-terms is PSPACE-hard. \qed
\end{theorem}

%
\begin{example}
Using the encoding where $\tau$ is set to  $\booltype_0$ in the types $\Delta_k^n$ for all $k,n\geq 0$, the formula $\forall x \exists y \exists z (x\zor y\zor z)\zand(\neg x\zor \neg y\zor \neg z)$ is represented by the safe term:
\begin{align*}
\sentail~&\mathbf{D}_0^2(\lambda x^{\booltype_2}. AND^0\\
&\quad\quad (\mathbf{D}_0^1(\lambda y^{\booltype_1}.OR^0\\
&\quad\quad\quad (\mathbf{D}_0^0(\lambda z^{\booltype_0}.OR^0\\
&\quad\quad\quad\quad (AND^0 (OR^0(OR^0~(\mathbf{C}_0^{2\mapsto 0}~x)~(\mathbf{C}_0^{1\mapsto 0}~y))z) \\
&\quad\quad\quad\quad\quad (OR^0(OR^0(NEG^0 (\mathbf{C}_0^{2\mapsto 0}~x))(NEG^0 (\mathbf{C}_0^{1\mapsto 0}~y)))(NEG^0~z))) \\
&\quad\quad\quad )F^0)\\
&\quad\quad)F^0)\\
&\quad) T^0 \\
& : \booltype_0 \enspace .
\end{align*}
\end{example}

\begin{remark}
The Boolean satisfaction problem (SAT) is just a particular instance of TQBF
where formulae are restricted to use only existential quantifiers, thus
the safe lambda calculus is also NP-hard. Asperti gave an interpretation of SAT in the simply-typed lambda calculus but his encoding relies on unsafe terms \cite{asperti-np}.
\end{remark}

\begin{remark}
\begin{enumerate}[(i)]
\item Because the safety condition restricts expressivity
in a non-trivial way, one can reasonably expect the
beta-eta equivalence problem to have a
lower complexity in the safe case than in the normal case; this intuition is strengthened by our failed attempt to encode type theory in the safe lambda calculus. No upper bounds is known at present. On the other hand our PSPACE-hardness result is probably a coarse lower bound; it would be interesting to know whether we also have EXPTIME-hardness.

\item 
Statman showed \cite{Statman:1979:TLE} that
when restricted to some finite set of types, the beta-eta equivalence problem is PSPACE-hard.
Such result is unlikely to hold in the safe lambda calculus. This is suggested by the fact that we had to use the entire type hierarchy to encode TQBF in the safe lambda calculus.
In fact we expect the beta-eta equivalence problem for safe terms to have a complexity lower than PSPACE when restricted to any finite set of types.

\item
The \emph{normalization problem} (``Given a (safe) term $M$, what is its $\beta$-normal form?'') is non-elementary. Indeed, let $\tau_{-2}
\equiv o$ and for $n\geq -1$, $\tau_n \equiv \tau_{n-1}\typear
\tau_{n-1}$. For $k,n\in \nat$, let $\overline{k}^n$ denote
the $k^{th}$ Church Numeral $\lambda s^{\tau_{n-1}}
z^{\tau_{n-2}} . s( \cdots (s(s\,z) \cdots)$ (with $k$ applications of $s$) of type $\tau_n$. Then for $n\geq1$, the safe term
$\overline{2}^{n-1}~\overline{2}^{n-2}\cdots \overline{2}^0$ of type
$\tau_0$ has size $\bigo(n)$ and its normal form
$\overline{\towerexp{n}{1}}^0$ has size $\bigo(\towerexp{n}{1})$.

Thus in the simply-typed lambda calculus, beta-eta equivalence is essentially as hard as normalization.
We do not know if this is the case in the safe lambda calculus.

\item 
A related problem is that of \emph{beta-reduction}: ``Given a $\beta$-normal term $M_1$
 and a term $M_2$, does $M_2$ $\beta$-reduce to $M_1$?''. It is known to be PSPACE-complete when restricted to order-$3$ terms \cite{schubert2001cbr}, but no complexity result is known for higher orders. The safe case can potentially give rise to interesting complexity characterizations at higher-orders.

\end{enumerate}
\end{remark}

\section{A game-semantic account of safety}
\label{sec:gamesemaccount} Our aim is to characterize safety by game
semantics. We shall assume that the reader is familiar with the
basics of game semantics; for an introduction, we recommend
Abramsky and McCusker's tutorial \cite{abramsky:game-semantics-tutorial}. Recall that a \emph{justified sequence} over an arena is an alternating sequence
of O-moves and P-moves such that every move $m$, except the opening
move, has a pointer to some earlier occurrence of the move $m_0$
such that $m_0$ enables $m$ in the arena. A \emph{play} is just a
justified sequence that satisfies Visibility and Well-Bracketing. A
basic result in game semantics is that $\lambda$-terms are denoted
by \emph{innocent strategies}, which are strategies that depend only
on the \emph{P-view} of a play. The main result
(Theorem~\ref{thm:safeincrejust}) of this section is that if a
$\lambda$-term is safe, then its game semantics (is an innocent
strategy that) is, what we call, \emph{P-incrementally justified}. In such a
strategy, pointers emanating from the P-moves of a play are uniquely
reconstructible from the underlying sequence of moves and pointers
from the O-moves therein: Specifically a P-question always points to
the last pending O-question (in the P-view) of a greater order.

The proof of Theorem~\ref{thm:safeincrejust} depends on a
Correspondence Theorem (see the Appendix) that relates the strategy
denotation of a $\lambda$-term $M$ to the set of \emph{traversals}
over a souped-up abstract syntax tree of the $\eta$-long form of $M$.
In the language of game semantics, traversals are just (concrete
representations of) the \emph{uncovering} (in the sense of Hyland
and Ong \cite{hylandong_pcf}) of plays in the strategy denotation.

The useful transference technique between plays and traversals was
originally introduced by the second author \cite{OngLics2006} for studying
the decidability of monadic second-order theories of infinite structures generated by
higher-order grammars (in which the $\Sigma$-constants or terminal symbols are at most
order 1, and \emph{uninterpreted}).
In the Appendix, we present an extension of this framework to the
general case of the simply-typed lambda calculus with free variables
of any order. A new traversal rule is introduced to handle nodes
labelled with free variables. Also new nodes are added to the
computation tree to account for the answer moves of the game
semantics, thus enabling the framework to model languages with
interpreted constants such as \pcf~(by adding traversal rules to
handle constant nodes).

\subsection*{Incrementally-bound computation tree}
In the context of higher-order grammars, the computation tree is
defined as the unravelling of the finite graph representing the \emph{long
transform} of the grammar \cite{OngLics2006}. Similarly we define the computation tree of
a $\lambda$-term as an abstract syntax tree of its $\eta$-long
normal form.  We write $l\langle t_1, \ldots, t_n \rangle$ with $n
\geq 0$ to denote the ordered tree with a root labelled $l$ with $n$
child-subtrees $t_1$, \ldots, $t_n$. In the following we consider arbitrary
simply-typed terms.

\begin{definition}\rm
\label{dfn:comptree}
  The \defname{computation tree} $\tau(M)$ of a simply-typed term
  $\Gamma \stentail M:T$ with variable names in a countable set
  $\mathcal{V}$ is a tree with labels in $$ \{ @ \} \union \mathcal{V}
  \union \{ \lambda x_1 \ldots x_n \ | \ x_1 ,\ldots, x_n \in
  \mathcal{V}, n\in\nat \}$$ defined from its $\eta$-long form as follows. Suppose $\overline{x} = x_1 \ldots x_n$ for $n\geq 0$ then
\begin{eqnarray*}
  \mbox{for $m\geq 0$, $z \in \mathcal{V}$: } \tau(\lambda {\overline x}^{\overline A} . z s_1 \ldots s_m : o) &=& \lambda {\overline x} \langle z \langle\tau(s_1),\ldots,\tau(s_m)\rangle\rangle \\
  \mbox{for $m \geq 1$: } \tau(\lambda {\overline x}^{\overline A} . (\lambda y^\tau.t) s_1 \ldots s_m :o) &=& \lambda \overline{x} \langle @ \langle \tau(\lambda y^\tau.t),\tau(s_1),\ldots,\tau(s_m) \rangle \rangle \ .
\end{eqnarray*}
\end{definition}

\begin{example}
\label{examp:comptree}
  Take $\stentail \lambda f^{o \typear o} .
(\lambda u^{o \typear o} . u) f : (o \typear o) \typear
o \typear o$.
\bigskip

\noindent
\begin{tabular}{cc}
Its $\eta$-long normal form is: & Its computation tree is:\\[8pt]
\begin{minipage}{0.45\textwidth}
\centering
$\begin{array}{ll}
 &\stentail  \lambda f^{o \typear o} z^o . \\
&\qquad(\lambda u^{o \typear o} v^o . u (\lambda.v)) \\
&\qquad(\lambda y^o. f y) \\
&\qquad(\lambda.z) \\
&: (o \typear o) \typear o \typear o
\end{array}$
\end{minipage}
&
\begin{minipage}{0.45\textwidth}
\centering
\psset{levelsep=5ex,linewidth=0.5pt,nodesep=1pt,arcangle=-20,arrowsize=2pt 1}
${\pstree{\TR{\lambda f z}}{\pstree{\TR{@}}{\pstree{\TR{\lambda u v}}{\pstree{\TR{u}}{\pstree{\TR{\lambda }}{\TR{v}}}}\pstree{\TR{\lambda y}}{\pstree{\TR{f}}{\pstree{\TR{\lambda }}{\TR{y}}}} \pstree{\TR{\lambda }}{\TR{z}}}}
}$
\end{minipage}
\end{tabular}
\end{example}

\begin{example}
  Take $\stentail \lambda u^o v^{((o \typear o) \typear o)} . (\lambda x^o . v (\lambda z^o . x)) u : o \typear ((o \typear o) \typear o) \typear o$.
  \bigskip

\noindent
\begin{tabular}{cc}
Its $\eta$-long normal form is: & Its computation tree is:\\[8pt]
\begin{minipage}{0.45\textwidth}
\centering
$\begin{array}{ll}
 &\stentail  \lambda u^o v^{((o \typear o) \typear o)} . \\
&\qquad(\lambda x^o . v (\lambda z^o . x)) u \\
&: o \typear ((o \typear o) \typear o) \typear o
\end{array}$
\end{minipage}
&
\begin{minipage}{0.45\textwidth}
\centering
$\pstree{\TR{\lambda u v}}{\pstree{\TR{@}}{\pstree{\TR{\lambda x}}{\pstree{\TR{v}}{\pstree{\TR{\lambda z}}{\TR{x}}}}\pstree{\TR{\lambda }}{\TR{u}}}}
$
\end{minipage}
\end{tabular}
\end{example}

Even-level nodes are $\lambda$-nodes (the root is on level 0). A
single $\lambda$-node can represent several consecutive variable
abstractions or it can just be a \emph{dummy lambda} if the
corresponding subterm is of ground type.  Odd-level nodes are
variable or application nodes.

The \defname{order} of a node $n$, written $\ord{n}$, is defined as
follows: @-nodes have order $0$. The order of a variable-node is the
type-order of the variable labelling it. The order of the root node
is the type-order of $(A_1,\ldots,A_p, T)$ where $A_1,\ldots, A_p$
are the types of the variables in the context $\Gamma$. Finally, the
order of a lambda node different from the root is the type-order of
the term represented by the sub-tree rooted at that node.

We say that a variable node $n$ labelled $x$ is \defname{bound} by a
node $m$, and $m$ is called the \defname{binder} of $n$, if $m$ is
the closest node in the path from $n$ to the root such that $m$ is
labelled $\lambda \overline{\xi}$ with $x\in \overline{\xi}$.

We introduce a class of computation trees in which the binder node
is uniquely determined by the nodes' orders:
\begin{definition}\rm
  A computation tree is \defname{incrementally-bound} if for all
  variable node $x$, either $x$ is \emph{bound} by the first
  $\lambda$-node in the path to the root with order $> \ord{x}$, or $x$
  is a \emph{free variable} and all the $\lambda$-nodes in the path to
  the root except the root have order $\leq \ord{x}$.
\end{definition}

\begin{proposition}[Safety and incremental-binding] \hfill
\label{prop:safe_imp_incrbound}
\begin{enumerate}[\em(i)]
\item If $M$ is safe then $\tau(M)$ is incrementally-bound.
\item Conversely, if $M$ is a \emph{closed} simply-typed term and $\tau(M)$
is incrementally-bound then $M$ is safe.
\end{enumerate}
\end{proposition}
\begin{proof}
  (i) Suppose that $M$ is safe. By Lemma
  \ref{prop:safe_iff_elnfsafe} the $\eta$-long form of $M$ is safe
  therefore $\tau(M)$ is the tree representation of a safe term.

In the safe lambda calculus, the variables in the context with the
lowest order must be all abstracted at once when using the
abstraction rule. Since the computation tree merges consecutive
abstractions into a single node, any variable $x$ occurring free in
the subtree rooted at a node $\lambda \overline{\xi}$ different from
the root must have order greater or equal to $\ord{\lambda
  \overline{\xi}}$. Conversely, if a lambda node $\lambda
\overline{\xi}$ binds a variable node $x$ then $\ord{\lambda
  \overline{\xi}} = 1+\max_{z\in\overline{\xi}} \ord{z} > \ord{x}$.

Let $x$ be a bound variable node. Its binder occurs in the path from
$x$ to the root, therefore, according to the previous observation,
$x$ must be bound by the first $\lambda$-node occurring in this path
with order $>\ord{x}$. Let $x$ be a free variable node then $x$ is
not bound by any of the $\lambda$-nodes occurring in the path to the
root. Once again, by the previous observation, all these
$\lambda$-nodes except the root have order smaller than $\ord{x}$.
Hence $\tau$ is incrementally-bound.

(ii) Let $M$ be a closed term such that $\tau(M)$ is
incrementally-bound.  W.l.o.g.~we can assume that $M$ is in $\eta$-long
form.  We prove that $M$ is safe by induction on its structure. The
base case $M \equiv \lambda \overline{\xi} . x$ for some variable $x$ is
trivial.  \emph{Step case:} If $M \equiv \lambda \overline{\xi} . N_1
\ldots N_p$.  Let $i$ range over $1..p$. We have $N_i \equiv \lambda
\overline{\eta_i} . N'_i$ for some non-abstraction term $N'_i$. By
the induction hypothesis, $\lambda \overline{\xi} . N_i = \lambda
\overline{\xi} \overline{\eta_i} . N'_i$ is a safe closed term, and
consequently $N'_i$ is necessarily safe. Let $z$ be a free variable
of $N'_i$ not bound by $\lambda \overline{\eta_i}$ in $N_i$. Since
$\tau(M)$ is incrementally-bound we have $\ord{z} \geq \ord{\lambda
  \overline{\eta_1}} = \ord{N_i}$, thus we can abstract the variables $\overline{\eta_1}$ using \rulenamet{abs} which shows that $N_i$ is safe.  Finally
we conclude $\sentail M = \lambda \overline{\xi} . N_1 \ldots N_p :
T$ using the rules \rulenamet{app} and \rulenamet{abs}.
\end{proof}

The assumption that $M$ is closed is necessary. For instance for
$x,y:o$, the computation trees $\tau(\lambda x y .x)$ and
$\tau(\lambda y . x)$ are both incrementally-bound but $\lambda x y
.x$ is safe and $\lambda y . x$ is not.

\subsection*{P-incrementally justified strategy}

We now consider the game-semantic model of the simply-typed lambda
calculus. The strategy denotation of a term-in-context $\Gamma
\stentail M : T$ is written $\sem{\Gamma
\stentail M : T}$. We define the \defname{order} of a move $m$,
written $\ord{m}$, to be the length of the path from $m$ to its
furthest leaf in the arena minus 1. (There are several ways to
define the order of a move; the definition chosen here is sound in
the current setting where each question move in the arena enables at
least one answer move.)

\begin{definition}\rm
  A strategy $\sigma$ is said to be \defname{P-incrementally
    justified} if for every play $s \, q \in \sigma$ where $q$ is a
  P-question, $q$ points to the last unanswered O-question in $\pview{s}$ with
  order strictly greater than $\ord{q}$.
\end{definition}
Note that although the pointer is determined by the P-view, the
choice of the move itself can be based on the whole history of the
play. Thus P-incremental justification does not imply innocence.

The definition suggests an algorithm that, given a play of a
P-incrementally justified denotation, uniquely recovers the pointers
from the underlying sequence of moves and from the pointers
associated to the O-moves therein. Hence:
\begin{lemma}
\label{lem:incrjustified_pointers_uniqu_recover} In P-incrementally
justified strategies, pointers emanating from P-moves are
superfluous.
\end{lemma}

\begin{example}
Copycat strategies, such as the identity strategy $id_A$ on game $A$
or the evaluation map $ev_{A,B}$ of type $(A \Rightarrow B) \times A
\typear B$, are all P-incrementally justified.\footnote{In such
strategies, a P-move $m$ is justified as follows: Either $m$ points
to the preceding move in the P-view or the preceding move is of
smaller order and $m$ is justified by the second last O-move in the
P-view.}
\end{example}

The Correspondence Theorem~\ref{thm:correspondence}
gives us the following equivalence:
\begin{proposition} 
\label{prop:Nher_incrbound_and_incrjustified} Let $\Gamma \stentail
M : T$ be a $\beta$-normal term. The computation tree $\tau(M)$ is
incrementally-bound if and only if $\sem{\Gamma \stentail M : T}$ is
P-incrementally justified.
\end{proposition}

\parpic[r]{
\pssetcomptree
\raisebox{-12pt}
{$\tree{\lambda^3}{\tree{f^2}{ \tree{\lambda y^1}{\TR{x^0} }}}$}
}
\begin{example}
Consider the $\beta$-normal term $\Gamma\stentail f (\lambda y .x) :
o$ where $y:o$ and $\Gamma =f:((o,o),o),~x:o$. The figure on the
right represents its computation tree with the node orders given as
superscripts.  The node $x$ is not incrementally-bound therefore $\tau(f
(\lambda y .x))$ is not incrementally-bound and by Proposition
\ref{prop:Nher_incrbound_and_incrjustified}, $\sem{\Gamma \stentail
f (\lambda y .x) : o}$ is not incrementally-justified (although
$\sem{\Gamma \stentail f : ((o,o),o)}$ and $\sem{\Gamma \stentail
\lambda
  y. x : (o,o)}$ are).
\end{example}
\smallskip

Propositions \ref{prop:safe_imp_incrbound} and
\ref{prop:Nher_incrbound_and_incrjustified} allow us to show the
following:
\begin{theorem}[Safety and P-incremental justification]
\label{thm:safeincrejust} \hfill
\begin{enumerate}[\em(i)]
\item If $\Gamma \sentail M : T$ then $\sem{\Gamma \sentail M : T}$ is P-incrementally justified.
\item If $\stentail M : T$ is a closed simply-typed term and $\sem{\stentail M : T}$ is P-incrementally justified then the $\beta$-normal form of $M$ is safe.
\end{enumerate}
\end{theorem}
\begin{proof} (i) Let $M$ be a safe simply-typed term. By Lemma
\ref{lem:safered_preserves_safety}, its $\beta$-normal form $M'$ is
also safe. By Proposition \ref{prop:safe_imp_incrbound}(i),
$\tau(M')$ is incrementally-bound and by Proposition
\ref{prop:Nher_incrbound_and_incrjustified}, $\sem{M'}$ is
incrementally-justified. Finally the soundness of the game model
gives $\sem{M} = \sem{M'}$.  (ii) is a consequence of Lemma
\ref{lem:safered_preserves_safety}, Proposition
\ref{prop:Nher_incrbound_and_incrjustified} and
\ref{prop:safe_imp_incrbound}(ii) and soundness of the game model.
\end{proof}

Putting Theorem \ref{thm:safeincrejust}(i) and Lemma
\ref{lem:incrjustified_pointers_uniqu_recover} together gives:
\begin{proposition}
  \label{prop:safe_ptr_recoverable} In the game semantics of safe
  $\lambda$-terms, pointers emanating from P-moves are unnecessary: they are uniquely recoverable from the underlying sequences of
  moves and from O-moves' pointers.
\end{proposition}

 \begin{example} If justification pointers are omitted then the denotations of the
   two Kierstead terms from Example~\ref{ex:kierstead} are not distinguishable.
   In the safe lambda calculus this ambiguity disappears
   since $M_1$ is safe whereas $M_2$ is not.
 \end{example}

In fact, as the last example highlights, pointers are superfluous at
order $3$ for safe terms whether from P-moves or O-moves. This is
because for question moves in the first two levels of an arena
(initial moves being at level $0$), the associated pointers are
uniquely recoverable thanks to the visibility condition. At the
third level, the question moves are all P-moves therefore their
associated pointers are uniquely recoverable by P-incremental
justification. This is not true anymore at order $4$: Take the safe
term-in-context $\psi:(((o^4,o^3),o^2),o^1) \sentail \psi (\lambda \varphi^{(o,o)} . \varphi a) : o^0$ for some constant $a:o$.
Its strategy denotation contains plays whose underlying sequence of
moves is $q_0 \, q_1 \, q_2 \, q_3 \, q_2 \, q_3 \, q_4$. Since
$q_4$ is an O-move, it is not constrained by P-incremental
justification and thus it can point to any of the two occurrences of
$q_3$.\footnote{More generally, a P-incrementally justified strategy
can contain plays that are not ``O-incrementally justified'' since
it must take into account any possible strategy incarnating its
context, including those that are not P-incrementally justified. For
instance in the given example, there is one version of the play that
is not O-incrementally justified (the one where $q_4$ points to the
first occurrence of $q_3$). This play is involved in the strategy
composition $\sem{ \stentail M_2 : (((o,o),o),o)} ; \sem{
\psi:(((o,o),o),o) \stentail \psi (\lambda \varphi . \varphi a):o}$
where $M_2$ denotes the unsafe Kierstead term.}

\subsection*{Towards a fully abstract game model}
The standard game models which have been shown to be fully abstract
for PCF \cite{abramsky94full,hylandong_pcf} are of course also fully
abstract for the restricted language safe PCF. One may ask, however,
whether there exists a fully abstract model with respect to safe
context only. Such model may be obtained by considering P-incrementally justified strategies---which have been shown to compose \cite{Blumphd}. Its is reasonable to think that O-moves also needs to be constrained by the symmetrical O-incremental justification, which corresponds to the requirement that contexts are safe. This line of work is still in progress.

\subsection*{Safe PCF and safe Idealised Algol}

\pcf\ is the simply-typed lambda calculus augmented with basic
arithmetic operators, if-then-else branching and a family of
recursion combinator $Y_A : ((A,A),A)$ for every type $A$.  We define
\emph{safe} \pcf\ to be \pcf\ where the application and abstraction
rules are constrained in the same way as the safe lambda calculus.
This language inherits the good properties of the safe lambda
calculus: No variable capture occurs when performing substitution
and safety is preserved by the reduction rules of the small-step
semantics of \pcf.

\subsubsection*{Correspondence}

The computation tree of a \pcf\ term is defined as the least
upper-bound of the chain of computation trees of its \emph{syntactic
approximants} \cite{abramsky:game-semantics-tutorial}.  It is
obtained by infinitely expanding the $\ycomb$ combinator, for instance
$\tau(Y (\lambda f x. f x))$ is the tree representation of the
$\eta$-long form of the infinite term $(\lambda f x. f x)
 ((\lambda f x. f x) ((\lambda f x. f x) ( \ldots$

It is straightforward to define the traversal rules modeling the
arithmetic constants of \pcf. Just as in the safe lambda calculus we
had to remove @-nodes in order to reveal the game-semantic
correspondence, in safe \pcf\ it is necessary to filter out the
constant nodes from the traversals. The Correspondence Theorem for
\pcf\ says that the revealed game semantics is isomorphic to the
set of traversals disposed of these superfluous nodes. This can
easily be shown for term approximants. It is then lifted to full
\pcf\ using the continuity of the function $\travset(\_)^{\filter
\theroot}$ from the set of computation trees (ordered by the
approximation ordering) to the set of sets of justified sequences of
nodes (ordered by subset inclusion). Finally computation trees of
safe \pcf\ terms are incrementally-bound thus we have
\begin{theorem}
\label{thm:safepcfpincr} Safe PCF terms have P-incrementally
justified denotations. \qed
\end{theorem}

Similarly, we can define safe \ialgol\ to be safe \pcf\ augmented
with the imperative features of Idealized Algol (\ialgol\ for short)
\cite{Reynolds81}.  Adapting the game-semantic correspondence and
safety characterization to \ialgol\ seems feasible although the
presence of the base type \iavar, whose game arena $\iacom^{\nat}
\times \iaexp$ has infinitely many initial moves, causes a mismatch
between the simple tree representation of the term and its game
arena. It may be possible to overcome this problem by replacing the
notion of computation tree by a ``computation directed acyclic
graph''.

The possibility of representing plays \emph{without some or all of
  their pointers} under the safety assumption suggests potential
applications in algorithmic game semantics. Ghica and McCusker
\cite{ghicamccusker00} were the first to observe that pointers are
unnecessary for representing plays in the game semantics of the
second-order finitary fragment of Idealized Algol ($\ialgol_2$ for
short). Consequently observational equivalence for this fragment can
be reduced to the problem of equivalence of regular expressions.  At
order $3$, although pointers are necessary, deciding observational
equivalence of $\ialgol_3$ is EXPTIME-complete
\cite{DBLP:journals/apal/Ong04,DBLP:conf/fossacs/MurawskiW05}.
Restricting the problem to the safe fragment of $\ialgol_3$ may lead
to a lower complexity.




\section{Further work and open problems}

The safe lambda calculus is still not well understood. Many basic
questions remain. What is a (categorical) model of the safe lambda
calculus? Does the calculus have interesting models?  What kind of
reasoning principles does the safe lambda calculus support, via the
Curry-Howard Isomorphism? Does the safe lambda calculus characterize
a complexity class, in the same way that the simply-typed lambda
calculus characterizes the polytime-computable numeric functions
\cite{DBLP:conf/tlca/LeivantM93}?  Is the addition of unsafe
contexts to safe ones conservative with respect to observational (or
contextual) equivalence?

With a view to algorithmic game semantics and its applications, it
would be interesting to identify sublanguages of Idealised Algol
whose game semantics enjoy the property that pointers in a play are
uniquely recoverable from the underlying sequence of moves. We name
this class PUR. $\ialgol_2$ is the paradigmatic example of a
PUR-language. Another example is \emph{Serially Re-entrant Idealized
  Algol} \cite{abramsky:mchecking_ia}, a version of \ialgol\ where
multiple uses of arguments are allowed only if they do not ``overlap
in time''.  We believe that a PUR language can be obtained by
imposing the \emph{safety condition} on $\ialgol_3$. Murawski
\cite{Murawski2003} has shown that observational equivalence for
$\ialgol_4$ is undecidable; is observational equivalence for
\emph{safe} $\ialgol_4$ decidable?

\subsection*{Acknowledgment} We thank Ugo
dal Lago for the insightful discussions we had during his visit at
the Oxford University Computing Laboratory in March 2008, and the anonymous referees for helpful comments.

    \bibliographystyle{abbrv} 
    \bibliography{../bib/dphil-all}

\begin{thebibliography}{10}

\bibitem{abramsky:mchecking_ia}
S.~Abramsky.
\newblock Semantics via game theory.
\newblock In {\em Marktoberdorf International Summer School}, 2001.
\newblock Lecture slides.

\bibitem{abramsky94full}
S.~Abramsky, P.~Malacaria, and R.~Jagadeesan.
\newblock Full abstraction for {PCF}.
\newblock In {\em Theoretical Aspects of Computer Software}, pages 1--15, 1994.

\bibitem{abramsky:game-semantics-tutorial}
S.~Abramsky and G.~McCusker.
\newblock Game semantics.
\newblock In H.~Schwichtenberg and U.~Berger, editors, {\em Logic and
  Computation: Proceedings of the 1997 Marktoberdorf Summer School}, pages
  1--56. Springer-Verlag, 1998.
\newblock Lecture notes.

\bibitem{safety-mirlong2004}
K.~Aehlig, J.~G. de~Miranda, and C.-H.~L. Ong.
\newblock Safety is not a restriction at level 2 for string languages.
\newblock Technical report, University of Oxford, 2004.

\bibitem{Aho68}
A.~V. Aho.
\newblock Indexed grammars -- an extension of context-free grammars.
\newblock {\em J. ACM}, 15(4):647--671, 1968.

\bibitem{asperti-np}
A.~Asperti.
\newblock P = {NP}, up to sharing.

\bibitem{Blumphd}
W.~Blum.
\newblock {\em The Safe Lambda Calculus}.
\newblock PhD thesis, University of Oxford, forthcoming.

\bibitem{blumong:safelambdacalculus}
W.~Blum and C.-H.~L. Ong.
\newblock The safe lambda calculus.
\newblock In S.~R.~D. Rocca, editor, {\em TLCA}, volume 4583 of {\em Lecture
  Notes in Computer Science}, pages 39--53. Springer, 2007.

\bibitem{Cau02}
D.~Caucal.
\newblock On infinite terms having a decidable monadic theory.
\newblock {\em Lecture Notes in Computer Science}, 2420:165--176, 2002.

\bibitem{Dam82}
W.~Damm.
\newblock The {IO-} and {OI}-hierarchy.
\newblock {\em TCS}, 20:95--207, 1982.

\bibitem{DG86}
W.~Damm and A.~Goerdt.
\newblock An automata-theoretical characterization of the {OI}-hierarchy.
\newblock {\em Information and Control}, 71(1-2):1--32, 1986.

\bibitem{demirandathesis}
J.~G. de~Miranda.
\newblock {\em Structures generated by higher-order grammars and the safety
  constraint}.
\newblock {D.P}hil thesis, University of Oxford, 2006.

\bibitem{DBLP:conf/sas/DimovskiGL05}
A.~Dimovski, D.~R. Ghica, and R.~Lazic.
\newblock Data-abstraction refinement: A game semantic approach.
\newblock In C.~Hankin and I.~Siveroni, editors, {\em SAS}, volume 3672 of {\em
  Lecture Notes in Computer Science}, pages 102--117. Springer, 2005.

\bibitem{DBLP:journals/jacm/FortuneLO83}
S.~Fortune, D.~Leivant, and M.~O'Donnell.
\newblock The expressiveness of simple and second-order type structures.
\newblock {\em J. ACM}, 30(1):151--185, 1983.

\bibitem{ghicamccusker00}
D.~R. Ghica and G.~McCusker.
\newblock Reasoning about idealized {\sc algol} using regular languages.
\newblock In {\em Proceedings of 27th International Colloquium on Automata,
  Languages and Programming ICALP 2000}, volume 1853 of {\em LNCS}, pages
  103--116. Springer-Verlag, 2000.

\bibitem{willgreenlandthesis}
W.~Greenland.
\newblock {\em Game Semantics for Region Analysis}.
\newblock PhD thesis, University of Oxford, 2004.

\bibitem{hmos-lics08}
M.~Hague, A.~S. Murawski, C.-H.~L. Ong, and O.~Serre.
\newblock Collapsible pushdown automata and recursive schemes.
\newblock {\em LICS}, pages 452--461, 2008.

\bibitem{hylandong_pcf}
J.~M.~E. Hyland and C.-H.~L. Ong.
\newblock On full abstraction for {PCF}: {I, II, and III}.
\newblock {\em Information and Computation}, 163(2):285--408, December 2000.

\bibitem{KNU02}
T.~Knapik, D.~Niwi{\'n}ski, and P.~Urzyczyn.
\newblock Higher-order pushdown trees are easy.
\newblock In {\em FOSSACS'02}, pages 205--222. Springer, 2002.
\newblock LNCS Vol.~2303.

\bibitem{DBLP:journals/tcs/Leivant93}
D.~Leivant.
\newblock Functions over free algebras definable in the simply typed lambda
  calculus.
\newblock {\em Theor. Comput. Sci.}, 121(1{\&}2):309--322, 1993.

\bibitem{DBLP:conf/tlca/LeivantM93}
D.~Leivant and J.-Y. Marion.
\newblock Lambda calculus characterizations of poly-time.
\newblock In M.~Bezem and J.~F. Groote, editors, {\em TLCA}, volume 664 of {\em
  Lecture Notes in Computer Science}, pages 274--288. Springer, 1993.

\bibitem{Loader1998}
R.~Loader.
\newblock Notes on simply typed lambda calculus, February 1998.

\bibitem{mairson1992spt}
H.~G. Mairson.
\newblock {A Simple Proof of a Theorem of Statman}.
\newblock {\em TCS}, 103(2):387--394, 1992.

\bibitem{Mas74}
A.~N. Maslov.
\newblock The hierarchy of indexed languages of an arbitrary level.
\newblock {\em Soviet Math. Dokl.}, 15:1170--1174, 1974.

\bibitem{Mas76}
A.~N. Maslov.
\newblock Multilevel stack automata.
\newblock {\em Problems of Information Transmission}, 12:38--43, 1976.

\bibitem{Meyer1974}
A.~R. Meyer.
\newblock The inherent computational complexity of theories of ordered sets.
\newblock In {\em Proc. Int'l. Cong. of Mathematicians}, volume~2, pages
  477--482, August 1974.

\bibitem{Murawski2003}
A.~S. Murawski.
\newblock On program equivalence in languages with ground-type references.
\newblock In {\em Logic in Computer Science, 2003. Proceedings. 18th Annual
  IEEE Symposium on}, pages 108--117, 22-25 June 2003.

\bibitem{DBLP:conf/fossacs/MurawskiW05}
A.~S. Murawski and I.~Walukiewicz.
\newblock Third-order idealized algol with iteration is decidable.
\newblock In V.~Sassone, editor, {\em FoSSaCS}, volume 3441 of {\em Lecture
  Notes in Computer Science}, pages 202--218. Springer, 2005.

\bibitem{DBLP:journals/apal/Ong04}
C.-H.~L. Ong.
\newblock An approach to deciding observational equivalence of algol-like
  languages.
\newblock {\em Ann. Pure Appl. Logic}, 130(1-3):125--171, 2004.

\bibitem{OngLics2006}
C.-H.~L. Ong.
\newblock On model-checking trees generated by higher-order recursion schemes.
\newblock In {\em Proceedings of IEEE Symposium on Logic in Computer Science.},
  pages 81--90. Computer Society Press, 2006.
\newblock Extended abstract.

\bibitem{OngHoMchecking2006}
C.-H.~L. Ong.
\newblock On model-checking trees generated by higher-order recursion schemes
  (technical report).
\newblock Preprint, 42 pp, 2006.

\bibitem{Reynolds81}
J.~C. Reynolds.
\newblock The essence of algol.
\newblock In J.~W. de~Bakker and J.~C. van Vliet, editors, {\em Algorithmic
  Languages}, pages 345--372. IFIP, North-Holland, Amsterdam, 1981.

\bibitem{schubert2001cbr}
A.~Schubert.
\newblock {The complexity of beta-reduction in low orders}.
\newblock {\em Proceedings TLCA 2001}, pages 400--414, 2001.

\bibitem{citeulike:622637}
H.~Schwichtenberg.
\newblock Definierbare funktionen im lambda-kalkul mit typen.
\newblock {\em Archiv Logik Grundlagenforsch}, 17:113--114, 1976.

\bibitem{Statman1979}
R.~Statman.
\newblock Intuitionistic propositional logic is polynomial-space complete.
\newblock {\em Theoretical Computer Science}, 9(1):67--72, July 1979.

\bibitem{Statman:1979:TLE}
R.~Statman.
\newblock The typed lambda-calculus is not elementary recursive.
\newblock {\em Theoretical Computer Science}, 9(1):73--81, July 1979.

\bibitem{DBLP:journals/tcs/Zaionc87}
M.~Zaionc.
\newblock Word operation definable in the typed lambda-calculus.
\newblock {\em Theor. Comput. Sci.}, 52:1--14, 1987.

\bibitem{DBLP:conf/aluacs/Zaionc88}
M.~Zaionc.
\newblock On the lambda-definable tree operations.
\newblock In C.~Bergman, R.~D. Maddux, and D.~Pigozzi, editors, {\em Algebraic
  Logic and Universal Algebra in Computer Science}, volume 425 of {\em Lecture
  Notes in Computer Science}, pages 279--292. Springer, 1988.

\bibitem{DBLP:journals/apal/Zaionc91}
M.~Zaionc.
\newblock Lambda-definability on free algebras.
\newblock {\em Ann. Pure Appl. Logic}, 51(3):279--300, 1991.

\bibitem{zaionc:csl94}
M.~Zaionc.
\newblock Lambda representation of operations between different term algebras.
\newblock {\em Lecture Notes in Computer Science}, pages 91--105, 1995.

\end{thebibliography}

    \section{Appendix -- Computation tree, traversals and correspondence}
    \label{sec:correspondence}
    The second author introduced the notion of
computation tree and traversals over a computation tree for the
purpose of studying trees generated by higher-order recursion
scheme \cite{OngLics2006}. Here we extend these concepts to the simply-typed lambda
calculus. Our setting allows the presence of free variables of any
order and the term studied is not required to be of ground type.
(This contrasts with \cite{OngLics2006}'s setting where the term is
of ground type and contains only \emph{uninterpreted
constant}.%
) Note that we automatically account for the
presence of uninterpreted constants since they can just be regarded
as free variables. We will then state the \emph{Correspondence
Theorem} (Theorem \ref{thm:correspondence}) that was used in Sec.
\ref{sec:gamesemaccount}.

In the following we fix a simply-typed term-in-context $\Gamma \stentail M :T$
(not necessarily safe) and we consider its computation tree
$\tau(M)$ as defined in Def.\ \ref{dfn:comptree}.

\subsection{Notations}
We first fix some notations. We write $\theroot$ to denote the root
of the computation tree $\tau(M)$. The set of nodes of this
computation tree is denoted by $\INodes$. The sets $\INodes_@$, $\INodes_\lambda$ and
$\INodes_{\sf var}$ are respectively the subset of @-nodes,
$\lambda$-nodes and variable nodes. The \defname{type} of a variable-labelled node is the
type of the variable that labels it; the type of the root is
$(A_1,\ldots,A_p, T)$ where $x_1:A_1,\ldots, x_p:A_p$ are the
variables in the context $\Gamma$; and the type of a node $n\in
(\INodes_\lambda \union \INodes_@) \setminus \{ \theroot \}$ is the type of the
subterm of $\elnf{M}$ corresponding to the subtree of $\tau(M)$ rooted at $n$.

\subsection{Pointers and justified sequences of nodes}

We define the \defname{enabling relation} on the set of nodes of the
computation tree as follows: $m$ enables $n$, written $m \enable n$,
if and only if $n$ is bound by $m$ (and we sometimes write $m
\enable_i n$ to indicate that $n$ is the $i^{\sf th}$ variable bound
by $m$); or $m$ is the root $\theroot$ and $n$ is a free variable;
or $n$ is a $\lambda$-node and $m$ is its parent node.

We say that a node $n_0$ of the computation tree is
\defname{hereditarily enabled} by $n_p \in \INodes$ if there are nodes
$n_1,\ldots, n_{p-1} \in \INodes$ such that $n_{i+1}$ enables $n_{i}$ for
all $i\in 0..p-1$.

For any set of nodes $S, H \subseteq N$ we write $S^{H\enable}$ for
$\{ n \in S \ | \exists m \in H \mbox{
s.t. } m  \enable^* n \}$ -- the subset of $S$ consisting of nodes
hereditarily enabled by some node in $H$. We will abbreviate
$S^{\{m\}\enable}$ into $S^{m\enable}$.

We call \defname{input-variables nodes} the elements of $\INodes_{\sf var}^{\theroot\enable}$ (\ie, variables that are hereditarily enabled by the root of $\tau(M)$). Thus we have $\INodes_{\sf var}^{\theroot\enable} = \INodes \setminus ( \INodes_{\sf var}^{\INodes_@\enable} \union \INodes_{\sf var}^{\INodes_\Sigma\enable})$.

A \defname{justified sequence of nodes} is a sequence of nodes with
pointers such that each occurrence of a variable or $\lambda$-node
$n$ different from the root has a pointer to some preceding
occurrence $m$ satisfying $m \enable n$. In particular, occurrences
of @-nodes do not have pointer. We represent the pointer in the
sequence as follows \Pstr[0.4cm]{ (m){m} \ldots (n-m,45:i) n }.
 where the label indicates that either $n$ is labelled with the $i^{th}$ variable
abstracted by the $\lambda$-node $m$ or that $n$ is the $i^{\sf th}$
child of $m$.  Children nodes are numbered from $1$ onward except for
@-nodes where it starts from $0$. Abstracted variables are numbered
from $1$ onward. The $i^{\sf th}$ child of $n$ is denoted by $n.i$.

We say that a node $n_0$ of a justified sequence is
\defname{hereditarily justified} by $n_p$ if there are occurrences $n_1,
\ldots, n_{p-1}$ in the sequence such that $n_i$ points to $n_{i+1}$
for all $i\in 0..p-1$. For any occurrence $n$ in a justified
sequence $s$, we write $s \filter n$ to denote the subsequence of
$s$ consisting of occurrences that are hereditarily justified by
$n$.

The notion of \defname{P-view} $\pview{t}$ of a justified sequence of
nodes $t$ is defined the same way as the P-view of a justified
sequences of moves in Game Semantics:\footnote{ The equalities in the
  definition determine pointers implicitly. For instance in the second
  clause, if in the left-hand side, $n$ points to some node in $s$
  that is also present in $\pview{s}$ then in the right-hand side, $n$
  points to that occurrence of the node in $\pview{s}$.}
$$\begin{array}{rclrcl}
\pview{\epsilon} &=&  \epsilon
& \pview{\Pstr[0.5cm]{ s \cdot (m) m \cdot \ldots \cdot (lmd-m,40){\lambda\overline{\xi}}
}}
 &=& \Pstr{
\pview{s} \cdot (m2) m \cdot (lm2-m2,50) {\lambda \overline{\xi}} } \\
\mbox{for $n \notin \INodes_\lambda$, } \pview{s \cdot n }  &=&  \pview{s} \cdot n \qquad
& \pview{s \cdot \theroot }  &=&  \theroot
\end{array}$$

The O-view of $s$, written $\oview{s}$, is defined dually. We will
borrow the game-semantic terminology: A justified sequences of nodes
satisfies \defname{alternation} if for any two consecutive nodes one
is a $\lambda$-node and the other is not, and \defname{P-visibility}
if every variable node points to a node occurring in the P-view a
that point.

\subsection{Computation tree with value-leaves}

We now add another ingredient to the computation tree that was not
originally used in the context of higher-order grammars \cite{OngLics2006}.  We write $\mathcal{D}$ to
denote the set of values of the base type $o$.  We add
\defname{value-leaves} to $\tau(M)$ as follows: For each value $v
\in \mathcal{D}$ and for each node of the computation tree we attach a new child
leaf $v_n$ to $n$. We write $\Nodes$ for the set of nodes (\ie, inner nodes and leaf nodes) of the
resulting tree. The set of leaf nodes is denoted $\LNodes$, we thus have $\Nodes = \INodes \union \LNodes$.
For $\$$ ranging in $\{@, \lambda, var \}$, we write $\Nodes_\$$ to denote the set consisting of nodes from $\INodes_\$$ together with leaf nodes with parent node in $\INodes_\$$; formally $\Nodes_\$ = \INodes_\$ \union \{ v_n \ | \ n \in \INodes_\$, v \in \mathcal{D} \}$.

The basic notions can be adapted to this new version of
computation tree: A value-leaf has order $0$. The enabling relation
$\enable$ is extended so that every leaf is enabled by its parent
node. A link going from a value-leaf $v_n$ to a node $n$ is labelled
by $v$ (\eg, \Pstr[0.4cm]{ (n) n \ldots (vn-n,35:v){v_n}}). For the
definition of P-view and visibility, value-leaves are treated as
$\lambda$-nodes if they are at an odd level in the computation tree,
and as variable nodes if they are at an even level.

We say that an occurrence of an inner node $n \in \INodes$ is
\defname{answered} by an occurrence $v_n$ if $v_n$ in
the sequence that points to $n$, otherwise we say that $n$ is
\defname{unanswered}. The last unanswered node is called the
\defname{pending node}.  A justified sequence of nodes is
\defname{well-bracketed} if each value-leaf occurring in it is justified by the pending node at that point.  If $t$ is a traversal then we write
$?(t)$ to denote the subsequence of $t$ consisting only of
unanswered nodes.

\subsection{Traversals of the computation tree}
\label{subsec:traversal}

A \emph{traversal} is a justified sequence of nodes of the computation tree where each node
indicates a step that is taken during the evaluation of the term.

\begin{definition}[Traversals for simply-typed $\lambda$-terms] \rm
\label{def:traversal} The set $\travset(M)$ of \defname{traversals}
over $\tau(M)$ is defined by induction over the rules of Table
\ref{tab:trav_rules}. A traversal that cannot be extended by any
rule is said to be \emph{maximal}.
\end{definition}
\begin{FramedTable}
\noindent {\bf Initialization rules}
\begin{itemize}[]
\item\rulenamet{Empty} $\epsilon \in \travset(M)$.
\item\rulenamet{Root} The sequence constituted of a single occurrence of $\tau(M)$'s root is a traversal.
\end{itemize}

\noindent {\bf Structural rules}
\begin{itemize}[]
    \item \rulenamet{Lam} If $t \cdot \lambda \overline{\xi}$ is a traversal then so is
        $t \cdot \lambda \overline{\xi} \cdot n$ where $n$
        denotes $\lambda \overline{\xi}$'s child and:
        \begin{compactitem}
            \item If $n \in \INodes_@ \union \INodes_\Sigma$ then it has no justifier;
            \item if  $n \in \INodes_{\sf var}\setminus \INodes_{\sf fv}$ then it points to the only occurrence\footnote{Prop.\ \ref{prop:pviewtrav_is_path} shows that P-views
            are paths in the tree thus $n$'s enabler occurs
            exactly once in the P-view.} of its binder in
            $\pview{t\cdot \lambda \overline{\xi}}$;
            \item if  $n \in \INodes_{\sf fv}$ then it points
            to the only occurrence of the root $\theroot$ in
            $\pview{t \cdot \lambda \overline{\xi}}$.
        \end{compactitem}
    \item \rulenamet{App} If $t \cdot @$ is a traversal then so is \Pstr[0.5cm]{t \cdot (m) @  \cdot (n-m,40:0) n}.
\end{itemize}

\emph{\bf Input-variable rules}
\begin{itemize}[]
\item \rulenamet{InputVar} If $t$ is a traversal where $t^\omega \in \INodes_{\sf var}^{\theroot\enable} \union \LNodes_\lambda^{\theroot\enable}$
and $x$ is an occurrence of a variable node in $\oview{t}$ then
so is $t \cdot n$ for every child $\lambda$-node $n$ of $x$, $n$
pointing to $x$.

\item \rulenamet{InputValue} If $t_1
\cdot x \cdot t_2$ is a traversal with pending node $x \in
\INodes_{\sf var}^{\theroot\enable}$ then so is \Pstr[0.5cm]{t_1 \cdot
(x){x} \cdot t_2 \cdot (xv-x,38:v){v_x} } for all $v \in
\mathcal{D}$.
\end{itemize}

\emph{\bf Copy-cat rules}
\begin{itemize}[]
\item\rulenamet{Var}
If \Pstr[0.5cm]{t \cdot (n){n} \cdot (lx){\lambda \overline{x}}
    \ldots (x-lx,50:i){x_i} } is a traversal where $x_i \in
    \INodes_{\sf var}^{@\enable}$ then so is \Pstr[0.5cm]{ t \cdot
(n){n} \cdot (lx){\lambda \overline{x}}  \ldots (x-lx,30:i){x_i}
    \cdot (letai-n,40:i){\lambda \overline{\eta_i}}
     }.

\item\rulenamet{Value}
  If \Pstr{t \cdot (m){m} \cdot (n){n}  \ldots
(vn-n,60:v){v}_{n} } is a traversal where $n\in \INodes$ then so is
\Pstr[0.7cm]{t \cdot (m){m} \cdot (n){n} \ldots
(vn-n,60:v){v}_{n} \cdot (vm-m,45:v){v}_m}.
\end{itemize}
\caption[Traversal rules for the simply-typed
lambda calculus]{Traversal rules for the simply-typed
$\lambda$-calculus.} \label{tab:trav_rules}
\end{FramedTable}

\parpic[r]{
 $\pssetcomptree\tree[levelsep=3ex,treesep=0.5cm]{\lambda} {
    \tree{@}{
        \pstree[linestyle=dotted]{\TR{\lambda y}\arclabel{0} }{
            \tree{y}{
                \tree{\lambda \overline{\eta_1}}{\vdots}
                \tree{\lambda \overline{\eta_i}}{\vdots}
                \tree{\lambda \overline{\eta_n}}{\vdots}
            }
        }
        \pstree[linestyle=dotted]{\TR{\lambda \overline{x}}
            \arclabel{1}}{ \tree{x_i}{\TR{} \TR{}}}
}}$ } A traversal always starts by visiting the root. Then it mainly
follows the structure of the tree. The (Var) rule permits us to jump
across the computation tree. The idea is that after visiting a
variable node $x$, a jump is allowed to the node corresponding to
the subterm that would be substituted for $x$ if all the
$\beta$-redexes occurring in the term were reduced. The sequence
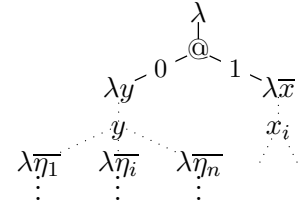
\Pstr[0.8cm]{\lambda \cdot (app) @  \cdot (ly) {\lambda y}  \ldots
(y-ly,35:1) y  \cdot (lx-app,38:1) {\lambda \overline{x}} \ldots
(x-lx,30:i) {x_i} \cdot (leta-y,40:i) {\lambda \overline{\eta_i} }
\ldots}
 is an example of traversal of the computation tree shown on the right.

\begin{example}
\label{examp:trav} The following justified sequence is a traversal
of the computation tree of example \ref{examp:comptree}:
$$\pstr[1.3cm]{\nd t= (n0){\lambda f z}
        \nd\cdot (n1){@}
        \nd\cdot (n2-n1){\lambda u v}
        \nd\cdot (n3-n2){u}
        \nd\cdot (n4-n1){\lambda y}
        \nd\cdot (n5-n0){f}
        \nd\cdot (n6-n5){\lambda }
        \nd\cdot (n7-n4){y}
        \nd\cdot (n8-n3){\lambda }
        \nd\cdot (n9-n2){v}
        \nd\cdot (n10-n1){\lambda }
        \nd\cdot (n11-n0){z} \ .
}$$
\end{example}

\begin{proposition} (Counterpart of the Path-traversal correspondence for higher-order grammars \cite[proposition 6]{OngHoMchecking2006}.)
\label{prop:pviewtrav_is_path}
Let $t$ be a traversal. Then:
\begin{enumerate}[\em(i)]
\item $t$ is a well-defined and well-bracketed justified sequence;
\item $t$ is a well-defined justified sequence satisfying alternation, P-visibility and O-visibility;
\item If $t$'s last node is not a value-leaf, then $\pview{t}$ is the path in the computation tree going from the root to $t$'s last node.
\end{enumerate}
\end{proposition}

The \defname{reduction} of a traversal $t$ is the subsequence of $t$
obtained by keeping only occurrences of nodes that are hereditarily
enabled by the root $\theroot$. This has the effect of eliminating
the ``internal nodes'' of the computation. If $t$ is a non-empty
traversal then the root $\theroot$ occurs exactly once in $t$ thus
the reduction of $t$ is equal to $t \filter r$ where $r$ is the
first occurrence in $t$ (the only occurrence of the root). We write
$\travset(M)^{\filter \theroot}$ for the set or reductions of
traversals of $M$.
\begin{example}
The reduction of the traversal given in example \ref{examp:trav} is:
$$ \Pstr[0.7cm]{t \filter \lambda f z = (n0){\lambda f z} \cdot (n1-n0){f}\cdot (n2-n1){\lambda }\cdot (n3-n0){z} } \ .$$
\end{example}

Application nodes are used to connect the operator and the operand
of an application in the computation tree but since they do not play
any role in the computation of the term, we can remove them from the
traversals.  We write $t-@$ for the sequence of nodes-with-pointers
obtained by removing from $t$ all @-nodes and value-leaves of
@-nodes, and where every pointer to an @-node is replaced by a pointer to its immediate predecessor in $t$. We write
$\travset(M)^{-@}$ for the set $\{ t - @ \ | \  t \in \travset(M)
\}$.
\begin{example}
Let $t$ be the traversal given in example \ref{examp:trav}, we have:
  $$\Pstr[1.3cm]{ t - @ = (n0){\lambda f z}\cdot (n1-n0){\lambda u v}\cdot (n2-n1){u}\cdot (n3-n0){\lambda y}\ (n4-n0){f}\cdot (n5-n4){\lambda }\cdot (n6-n3){y}\cdot (n7-n2){\lambda }\cdot (n8-n1){v}\cdot (n9-n0){\lambda }\cdot (n10-n0){z}} \ .$$
\end{example}

\begin{remark}
Clearly if $M$ is $\beta$-normal then $\tau(M)$ does not contain any
@-node therefore all nodes are hereditarily enabled by the root and
we have $\travset(M)^{-@} = \travset(M) = \travset(M)^{\filter
\theroot }$.
\end{remark}

\begin{lemma}
\label{lem:betanf_trav_pview_red} Suppose that $M$ is a
$\beta$-normal simply-typed term. Let $t$ be a non-empty traversal
of $M$ and $r$ denote the only occurrence of $\tau(M)$'s root in
$t$. If $t$'s last occurrence is not a leaf then
$$ \pview{t} \filter r = \pview{?(t) \filter  r }\ .$$
\end{lemma}
In the lambda calculus without interpreted constants this lemma
follows immediately from the fact that $\travset(M) =
\travset(M)^{\filter \theroot}$. It remains valid in the
presence of interpreted constants provided that the traversal rules
implementing the constants are \emph{well-behaved}\footnote{A
traversal rule is \defname{well-behaved} if it can be stated under
the form ``$t = t_1\cdot n \cdot t_2 \in \travset(M)\ \zand\ ?(t) =
?(t_1) \cdot n \zand n \in \INodes_{\Sigma}\union \INodes_{\sf var} \zand P(t)
\zand m\in S(t) \imp\ $\Pstr{ t_1\cdot (n){n} \cdot t_2 \cdot
(m-n,25){m} \in {\travset(M)}}'' for some expression $P$ expressing
a condition on $t$ and function $S$ mapping traversals of the form
of $t$ to a subset of the children of $n$.}.

\subsection{Computation trees and arenas}
We consider the well-bracketed game model of the simply-typed lambda
calculus.  We choose to represent strategies using ``prefix-closed
set of plays''.\footnote{In the literature, a strategy is commonly
defined as a set of plays closed by taking a prefix of \emph{even}
length. However for the purpose of showing the correspondence with
traversals, the ``prefix-closed''-based definition is more
adequate.} We fix a term $\Gamma \stentail M : T$ and write
$\sem{\Gamma \stentail M : T}$ for its strategy denotation. The
answer moves of a question $q$ are written $v_q$ where $v$ ranges in
$\mathcal{D}$.

\begin{proposition}
There exists a function $\varphi_M$, constructible from $M$,
that maps nodes from $\Nodes\setminus (\Nodes_@ \union \Nodes_\Sigma)$ to moves of
the arenas underlying the strategy denotations of $M$'s subterms such that:
\begin{enumerate}[$\bullet$]
\item $\varphi$ maps $\lambda$-nodes to O-questions, variable
nodes to P-questions, value-leaves of $\lambda$-nodes to
P-answers and value-leaves of variable nodes to O-answers.

\item $\varphi$ maps nodes of a given order to moves of
the same order.
\end{enumerate}
\end{proposition}
If $t = t_0 t_1 \ldots$ is a justified sequence of nodes in
$\Nodes_\lambda \union \Nodes_{\sf var}$ then $\varphi(t)$ is defined to
be the sequence of moves $\varphi(t_0)\ \varphi(t_1) \ldots$
equipped with the pointers of $t$.

\begin{example}
Take $\lambda x . (\lambda g . g x) (\lambda y . y)$ with $x,y:o$ and $g:(o,o)$.
The diagram below represents the computation tree (middle), the arenas
$\sem{(o,o), o}$ (left), $\sem{o , o}$ (right), $\sem{o\rightarrow o}$ (rightmost)
and $\varphi = \psi \union \psi_{\lambda g.g x}^{\lambda g, q_{\lambda g}} \union
\psi_{\lambda y.y}^{\lambda y, q_{\lambda y}}$
(dashed-lines).
$$\psset{levelsep=3.5ex}
\pstree{\TR[name=root]{\lambda x}}
{
    \pstree{\TR[name=App]{@}}
    {
            \pstree{\TR[name=lg]{\lambda g}}
                { \pstree{\TR[name=lgg]{g}}{
                        \pstree{\TR[name=lgg1]{\lambda}}
                        { \TR[name=lgg1x]{x}  } } }
            \pstree{\TR[name=ly]{\lambda y}}
                    {\TR[name=lyy]{y}}
    }
}
\rput(4.5cm,-1cm){
  \pstree{\TR[name=A1lx]{q_{\lambda x}}}
        { \TR[name=A1x]{q_x} }
}
\rput(-6cm,-1.5cm){
    \pstree{\TR[name=A2lg]{q_{\lambda g}}}
    {
        \pstree{\TR[name=A2g]{q_g}}
        {  \TR[name=A2g1]{q_{g_1}}   }
    }}
\rput(2.5cm,-1.5cm){
    \pstree{\TR[name=A3ly]{q_{\lambda y}}}
        { \TR[name=A3y]{q_y}
        }
}
\psset{nodesep=1pt,arrows=->,arcangle=-20,arrowsize=2pt 1,linestyle=dashed,linewidth=0.3pt}
\ncline{->}{root}{A1lx} \mput*{\psi}
\ncarc{->}{lgg1x}{A1x}
\ncline{->}{lg}{A2lg} \mput*{\psi_{\lambda g.g x}^{\lambda g, q_{\lambda g}}}
\ncline{->}{lgg}{A2g}
\ncline{->}{lgg1}{A2g1}
\ncline{->}{ly}{A3ly} \mput*{\psi_{\lambda y.y}^{\lambda y, q_{\lambda y}}}
\ncline{->}{lyy}{A3y}
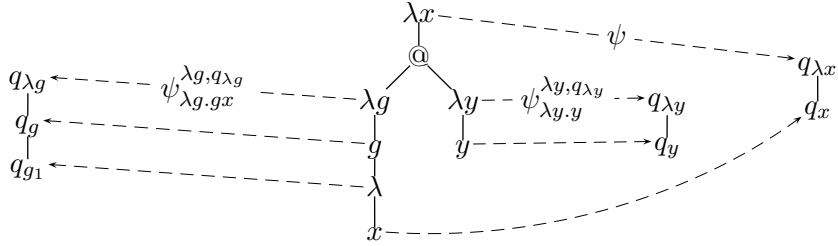
$$
\end{example}

\subsection{The Correspondence Theorem}

In game semantics, strategy composition is performed using a
CSP-like ``composition + hiding''. If some of the internal moves are
not hidden then we obtain alternative denotations called
\defname{revealed semantics} \cite{willgreenlandthesis} or
\emph{interaction} semantics \cite{DBLP:conf/sas/DimovskiGL05}.
We obtain different notions of revealed semantics depending on the
choice of internal moves that we hide. For instance the
\defname{fully revealed denotation} of $\Gamma \stentail M
:T$, written $\revsem{\Gamma \stentail M : T}$, is obtained by
uncovering all the internal moves from $\sem{\Gamma \stentail M :
T}$ that are generated during composition.\footnote{An algorithm
that uniquely recovers hidden moves from $\sem{\Gamma \stentail M :
T}$ was given by Hyland and Ong \cite[Part II]{hylandong_pcf}.} The inverse operation consists in filtering out the internal moves.

The \defname{syntactically-revealed denotation}, written
$\syntrevsem{\Gamma \stentail M : T}$, differs from the
fully-revealed one in that only certain internal moves are preserved
during composition: When computing the denotation of an application
joint by an @-node in the computation tree, all the internal moves
are preserved. When computing the denotation of $\revsem{y_i N_1
\ldots N_p}$ for some variable $y_i$, however, we only preserve the
internal moves of $N_1$, \ldots, $N_p$ while omitting the internal
moves produced by the copy-cat projection strategy denoting $y_i$.

The Correspondence Theorem states that in the simply-typed lambda calculus, the set $\travset(M)$ of traversals of the computation tree is isomorphic to the
syntactically-revealed denotation, and the set of traversal
reductions is isomorphic to the standard strategy denotation:
\begin{theorem}[The Correspondence Theorem]
\label{thm:correspondence} We have the following two isomorphisms:
\begin{eqnarray*}
(i)~\varphi_M  &: \travset(M)^{-@} \stackrel{\cong}{\longrightarrow} \syntrevsem{\Gamma \stentail M :T} \\
(ii)~\varphi_M  &: \travset(M )^{\filter \theroot} \stackrel{\cong}{\longrightarrow} \sem{\Gamma \stentail M : T} \ .
\end{eqnarray*}
\end{theorem}
\bigskip

\begin{example}
Take the term $M \equiv \lambda f^{(o,o)} z^o . (\lambda g^{(o,o)} x . f x) (\lambda y^o. y) (f z)$
of type $((o,o),o, o)$.  The figure below represents the computation tree
(left tree), the arena $\sem{((o,o),o, o)}$ (right tree) and
the function $\psi_M$ (dashed line). (Answer moves are not shown for clarity.)
Take the traversal $t$ given hereunder, we have:
\bigskip

\noindent\begin{tabular}{cc}
\begin{minipage}{6cm}
\centering
$\pssetcomptree\pstree[levelsep=2.5ex,treesep=0.3cm]{ \TR[name=root]{\lambda f z} }
     {  \tree[levelsep=4ex]{@}
        {   \tree{\lambda g x}{
                  \pstree{\TR[name=f]{f^{[1]}}}{
                            \pstree{\TR[name=lmd]{\lambda^{[2]}}}
                                {\TR{x}}
                  }
                }
            \tree{\lambda y }{\TR{y}}
            \tree{\lambda ^{[3]}}{
                \pstree{\TR[name=f2]{f^{[4]}}} {
                \pstree{\TR[name=lmd2]{\lambda^{[5]}}}{\TR[name=z]{z}}
                }
            }
        }
     }
\hspace{2cm}
  \pstree[levelsep=8ex, treesep=0.3cm]{ \TR[name=q0]{q^0} }
    {   \pstree[levelsep=4ex]{\TR[name=q1]{q^1}} {\TR[name=q2]{q^2}}
        \TR[name=q3]{q^3}
    }
\psset{nodesep=1pt,arrows=->,arrowsize=2pt 1,linestyle=dashed,linewidth=0.3pt}
\ncline{->}{root}{q0} \mput*{\psi_M}
\ncarc[arcangle=-25]{->}{z}{q3}
\ncarc[arcangle=10]{->}{f}{q1}
\ncarc[arcangle=10]{->}{lmd}{q2}
\ncline{->}{f2}{q1}
\ncline{->}{lmd2}{q2}$
\end{minipage}
&
\begin{minipage}{8cm}
\begin{asparablank}
  \item  \Pstr[1cm]{
t = (n){\lambda f z} \cdot
(n2){@} \cdot
(n3-n2,60){\lambda g x} \cdot
(n4-n,45){f^{[1]}}\cdot
(n5-n4,45){\lambda^{[2]}} \cdot
(n6-n3,45){x} \cdot
(n7-n2,35){\lambda^{[3]}} \cdot
(n8-n,35){f^{[4]}} \cdot
(n9-n8,45){\lambda^{[5]}} \cdot
(n10-n,35){z}
}

\item \Pstr[1.2cm]{
t\filter r = (n){\lambda f z} \cdot (n4-n,50){f}^{[1]} \cdot
(n5-n4,60){\lambda}^{[2]} \cdot (n8-n,45){f}^{[4]} \cdot
(n9-n8,60){\lambda}^{[5]} \cdot (n10-n,40){z}}
\item
\Pstr[1.2cm]{ {\varphi_M(t\filter r) =\ } (n){q^0}\cdot
(n4-n,60){q^1}\cdot (n5-n4,60){q^2}\cdot (n8-n,45){q^1}\cdot
(n9-n8,60){q^2}\cdot (n10-n,38){q^3} \in \sem{M}\ .}
\end{asparablank}
\end{minipage}
\end{tabular}
\end{example}

\end{document}